\documentclass[final,3p,sort&compress,times]{elsarticle}
\usepackage{bm}
\usepackage{amsmath}
\usepackage{amssymb}
\usepackage{gensymb}        % To introduce degree celcius
\usepackage{caption}        % To introduce degree celcius
\usepackage{hyperref}       % Use package to create hyperlinks to equations and sections 
\usepackage{soul}           % Package to strike through text
\usepackage{dcolumn}        % Align table columns on decimal point
\usepackage{graphicx}       % Include figure files
\graphicspath{{Pictures/}}  % Include path to figures 
\usepackage{subfiles}       % subfiles package  to use additional .tex subfiles as sections 
\usepackage{subfig}
\usepackage{xcolor,soul}
\sethlcolor{lightgray}
\usepackage[utf8]{inputenc}
\usepackage[T1]{fontenc}
\usepackage{epstopdf}
\usepackage{placeins}
\usepackage{chngcntr}

\newcommand{\tildbm}[1]{\tilde{\bm{#1}}}
\newcommand{\EMPS}{\textit{EMPS} }
\newcommand{\PDMS}{\textit{PDMS} }

\newcommand{\tr}{\text{tr}}

\journal{Journal XXX}

\begin{document}

\begin{frontmatter}

    \title{Phase-field modeling of elastic microphase separation}

    \author[inst1]{H. Oudich}
    \author[inst1]{P. Carrara}
    \author[inst1]{L. De Lorenzis\corref{cor1}}
    \affiliation[inst1]{organization={Eidgenössische Technische Hochschule (ETH) Zürich, Computational Mechanics Group},
    addressline={Tannenstrasse 3}, 
    city={Zürich},
    postcode={8092}, 
    country={Switzerland}}
    
    \cortext[cor1]{Corresponding author}
    \ead{ldelorenzis@ethz.ch}

    %%%%%% Abstract section  and keywords %%%%%%%%%%%%%%%%%%%%%%%%%%%%%%%%%%%%%%%%
    \begin{abstract}
        \label{abstract}

We propose a novel phase-field model to predict elastic microphase separation in polymer gels. 
To this end, we extend the \textit{Cahn-Hilliard} free-energy functional to incorporate an elastic strain energy  and a coupling term. These contributions are naturally obtained from a derivation that starts from an  entropic elastic energy density combined with the assumption of weak compressibility, upon second-order approximation around the swollen state. The resulting  terms correspond to those of a poroelastic formulation where the coupling energetic term can be interpreted as the osmotic work of the solvent within the polymer matrix.
Additionally, a convolution term is included in the total energy to model non-local forces responsible for coarsening arrest. With analytical derivations in 1D and finite element computations in 2D we show that the mechanical deformation controls the composition of the stable phases, the initial characteristic length and time, the coarsening rates and the arrested characteristic length. Moreover, we demonstrate that the proposed coupling is able to predict the arrest of coarsening at a length scale controlled by the stiffness of the dry polymer. The numerical results show excellent agreement with the experimental evidence in terms of phase-separated morphology and scaling of the characteristic length with the stiffness of the dry polymer.

    \end{abstract}
    
    
\end{frontmatter}
%%%%%%%%%%%%%%%%%%%%%%%%%%%%%%%%%%%%%%%%%%%%%%%%%%%%%%%

%%%%%%%%%%%%%%%%%%%%%%%%%%%%%%%%%%%%%%%%%%%%%%%%%%%%%%%
%%%%%%%%%%%%%%%%%%%%% main text %%%%%%%%%%%%%%%%%%%%%%%
%%%%%%%%%%%%%%%%%%%%%%%%%%%%%%%%%%%%%%%%%%%%%%%%%%%%%%%

%%%%%% Introduction section %%%%%%%%%%%%%%%%%%%%%%%%%%%%%%%%%%%%%%%%
\section{Introduction}  
\label{sec:Introduction}

Spinodal patterns are two-phase, channel-like architectures that characterize the microstructure of different natural materials and are responsible for specific aspects of their behavior, such as the high strength of the skeleton of sea urchins \cite{yang_high_2022}, the improved hardening of {Al/Zn} alloys \cite{douglass_spinodal_1969, findik_improvements_2012} or the photonic properties of the feathers of some birds \cite{dufresne_self-assembly_2009} among others \cite{brangwynne_germline_2009, gibaud_closer_2009, fernandez-rico_putting_2022}. 
Spinodal microstructures naturally arise from an initially homogeneous mixture which becomes thermodynamically unstable as a result of external stimuli, driving the system to reach a lower energy state. This usually involves the spontaneous separation of the mixture in two phases, each with a specific composition, followed by the emergence of the characteristic intertwined channel-like spinodal structures \cite{wodo_computationally_2011}. The transition from homogeneous to phase-separated state is termed \textit{spinodal decomposition} and can be subdivided in two main stages: an \textit{early stage} where the homogeneous mixture phase separates, forming a pattern of alternating clusters of the two stable compositions separated by interfaces, and a \textit{coarsening stage} where the clusters expand in size and reduce in number. Diffusion drives the motion of the species and leads clusters with the same composition to merge at the expenses of the number of interfaces. The spinodal pattern emerging from the early  stage exhibits a characteristic length scale, which steadily increases during coarsening following the so-called \textit{Ostwald} ripening law \cite{ostwald_studien_1897, voorhees_ostwald_nodate}. The system evolves until the two stable phases are separated by a single interface and hence collected in two large clusters with size fulfilling the mass balance condition. However, if the coarsening phase is arrested, the final microstructure has specific characteristic length and compositions of the phases, which confer to the material specific properties. 

Inducing spinodal decomposition in initially homogeneous mixtures is possible relying on different stimuli such as irradiation \cite{Lu2021}, pressure change \cite{Liu2001} or temperature quenching \cite{fernandez-rico_putting_2022}. More challenging is to  arrest the coarsening stage at a specific characteristic length, so as to control the topology and properties of the final microstructure. To this end, different techniques are available, e.g. vitrification (gelation), solvent evaporation, photopolymerization and the use of di-block copolymers \cite{manley_glasslike_2005, gao_microdynamics_2015, ge_evidence_2000, tran-cong-miyata_phase_2017, leibler_theory_1980, thomas_ordered_1986}. However, they lead to a limited possibility to control the characteristic length and an imperfect uniformity of the final pattern \cite{fernandez-rico_putting_2022, fernandez-rico_elastic_2023}. A more recent approach termed \textit{elastic microphase separation} (\textit{EMPS}) leverages the elastic properties of rubbery polymers to arrest coarsening and the first experimental tests seem to confirm its higher efficiency and flexibility compared to the aforementioned approaches \cite{fernandez-rico_elastic_2023}.

\begin{figure}[!hbt]
    \centering
    \captionsetup{justification=centering}
    \includegraphics[scale = 0.5]{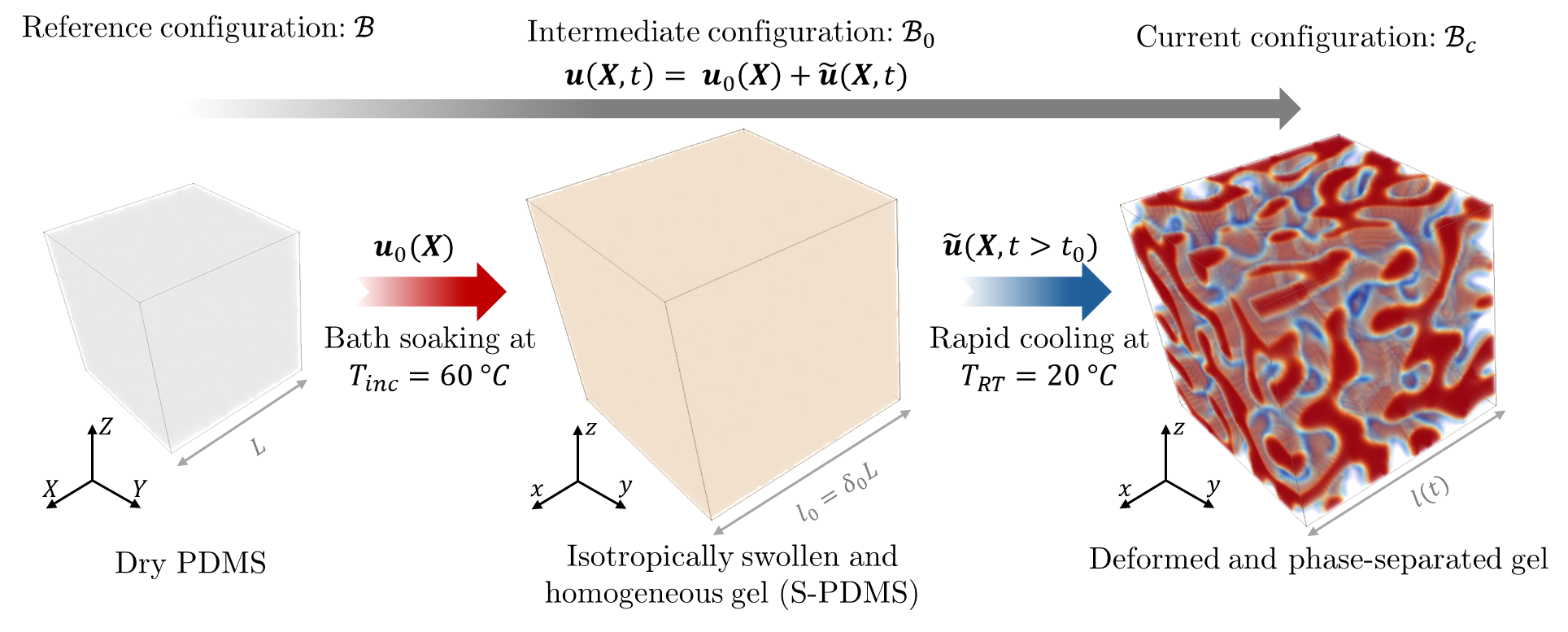}
    \caption{Scheme of EMPS \cite{fernandez-rico_elastic_2023}.}
\label{fig:EMPS_process_scheme}
\end{figure}

EMPS, in its instance reported in \cite{fernandez-rico_elastic_2023}, is schematized in Fig.~\ref{fig:EMPS_process_scheme} and involves soaking an initially dry specimen of poly-dimethyl-siloxane (\textit{PDMS}) (also referred to as dry polymer) in a bath of hepta-fluoro-butyl methacrylate (\textit{HFBMA}) oil (also denoted as solvent) at an incubation temperature of $T_{inc}=60 \degree C$ until saturation. During this phase the PDMS swells due to oil intake resulting in a homogeneous and isotropically swollen gel, which we denote as \textit{S-PDMS}. The gel is then rapidly cooled (i.e., quenched) to room temperature $T_{R}=20 \degree C$, triggering spinodal decomposition.  Bright-field optical microscopy analyses reveal the presence of a spinodal microstructure, whereby the two phases exhibit small differences in terms of the local volume fractions of oil compared to the initial homogeneous mixture. Also, the experimental evidence shows that the characteristic length of the spinodal microstructure does not change over time, i.e. coarsening is arrested. In \cite{fernandez-rico_elastic_2023}, it is reported that phase separation and arrest of coarsening took place rapidly after quenching, therefore, it was not possible to observe the early stages of the spinodal decomposition nor the potential occurrence of a partial coarsening. Hence, it remains unclear whether the measured stabilized characteristic lengths coincide with the initial characteristic lengths or are larger due to partial coarsening. Here we refer to the initial dry PDMS state as \textit{reference} configuration ($\mathcal{B}$), while the S-PDMS gel at incipient phase separation (i.e., after quenching) and the phase-separated system are referred to as \textit{intermediate} ($\mathcal{B}_0$) and \textit{current} ($\mathcal{B}_c$) configurations, respectively (Fig.~\ref{fig:EMPS_process_scheme}).

The main idea of EMPS is to exploit the competition between chemical de-mixing driving forces and elastic deformations arising in the polymer network as a consequence of local changes in composition triggered by phase separation. To investigate this competition, in \cite{fernandez-rico_elastic_2023} experiments are conducted with different PDMS materials varying the amount of crosslinker and therefore the Young's modulus of the dry PDMS, $E_{PDMS}$. For PDMS stiffnesses $E_{PDMS}= 180, \, 350, \,  800 \, kPa$, the measured characteristic length, $\ell_0$, is found to scale inversely with the square root of $E_{PDMS}$, i.e. $\ell_0 \propto 1/\sqrt{E_{PDMS}}$. Tab.~\ref{tab:table_PDMS_exp_data} collects the experimental data that we later use for the calibration of the model parameters. Here $\varphi_0$ is the volume fraction of the solvent in the swollen configuration, $s_0$ is the swelling ratio, i.e. the volume increase relative to the volume of the dry polymer, and $E_0$ is the \textit{Young's} modulus of the homogeneous S-PDMS mixture.
\begin{table}[]
    \centering
    \begin{tabular}{ccccccc}
$E_{PDMS}$ & $\varphi_0$ & $s_0$ & $E_0$    & $\ell_0$  \\
$(kPa)$    & $(-)$       & $($\%$)$           & $(kPa)$  & $(\mu m)$ \\
$180$      & $0.65$      & $44.0$          & $35.2$  & $2.0$     \\
$350$      & $0.62$      & $43.8$          & $69.2$  & $1.6$     \\
$800$      & $0.56$      & $41.2$          & $157.2$ & $0.9$      
    \end{tabular}
    \caption{\label{tab:table_PDMS_exp_data}Summary of experimental results for three different PDMS types.}
\end{table}

When it comes to modeling, while the role of elasticity in spinodal decomposition is acknowledged since the early studies on the topic \cite{cahn_spinodal_1961}, to the best of our knowledge a model able to reproduce EMPS while quantitatively capturing the influence of the stiffness of the dry polymer is not yet available, see Section \ref{sec:Literature_review} for a literature review.
Thus, the aim of the present study is to formulate a model coupling spinodal decomposition and mechanics able to reproduce the main experimental evidence of \cite{fernandez-rico_elastic_2023}, first and foremost the effect of elasticity on characteristic length scales and coarsening arrest. To this end, we depart from the coupled theory of diffusion and large deformation in polymeric gels by \cite{hong_theory_2008}, based on the Flory-Rehner theory, linearized by \cite{bouklas_swelling_2012, bouklas_nonlinear_2015} around the intermediate (swollen) configuration and shown to be consistent with Biot's linear poroelasticity theory to describe small deformations in an isotropically swollen gel \cite{biot_general_1941}. We then augment this theory by enabling it to reproduce spinodal decomposition in a similar fashion as in  the  \textit{Cahn-Hilliard} equation \cite{cahn_free_nodate}, which leads to a phase-field model of EMPS. Finally, we introduce an additional free energy density term, inspired by the \textit{Cahn-Oono} equation \cite{ohta_equilibrium_1986}, to enable the prediction and calibration of coarsening arrest. 

The paper is structured as follows. Section \ref{sec:Literature_review} provides a concise literature review on models of spinodal decomposition, coupled or not with mechanics, and highlights their shortcomings in predicting EMPS. Section \ref{app:derivation_coupled_model} formulates the free energy density of our novel phase-field model, which is the basis for the derivation of its governing equations.
These equations are reported in Section \ref{sec:1d_model} for the special case of the 1D setting, which is unrealistic but allows us to derive analytical results using linear stability analysis and obtain some useful insights. The governing equations in the multi-dimensional setting are presented in Section \ref{sec:2d_model} and comparisons with the experimental evidence in \cite{fernandez-rico_elastic_2023} are performed. Finally, the main conclusions are summarized in Section \ref{sec:conclusions}.

%%%%%% Literature review section %%%%%%%%%%%%%%%%%%%%%%%%%%%%%%%%%%%
\section{Brief literature review on modeling of spinodal decomposition}  
\label{sec:Literature_review}

%The uncoupled spinodal decomposition model: Cahn-Hilliard 
A milestone in the modeling of spinodal decomposition is represented by the  {Cahn-Hilliard} equation \cite{cahn_free_nodate}, which describes the behavior of a mixture where the chemical species diffuse and phase separate. The local composition of the material is indicated by a conserved parameter termed \textit{phase-field variable} $\phi$ and the governing equations are derived by minimizing the energy of the system under the mass conservation constraint. The free energy density reads

\begin{equation}
    \label{eq:ch_en_generic}
    \psi_{CH}(\phi,\nabla\phi) = \psi_{GL}(\phi) + \psi_{int}(\nabla\phi)\,,
\end{equation}
\noindent where $\psi_{GL}(\phi)$ is the \textit{Ginzburg-Landau} chemical energy density and $\psi_{int}(\nabla\phi)$ is the regularized interfacial energy density depending on the gradient of the phase-field variable $\nabla\phi$ \cite{provatas_phase-field_2010}. 

Numerous studies demonstrate that the Cahn-Hilliard equation is able to qualitatively reproduce the main stages of spinodal decomposition, including the early phase separation and the late coarsening stage \cite{novick-cohen_nonlinear_1984, chan_computational_1995, elder_early_1988, rogers_numerical_1989, kuhl_computational_2007, novick-cohen_chapter_2008, agosti_cahn-hilliardtype_2017, konig_two-dimensional_2021, yang_phase_2023}. However, when it comes to quantitative predictions, the {Cahn-Hilliard} model also has limitations. According to \cite{cabral_spinodal_2018}, reproducing the experimental early-stage characteristic length requires the adjustment of the model parameters for each investigated mixture or concentration. Additionally, the model parameters needed to reproduce the early-stage behavior are usually not suitable to predict the coarsening stage. Therefore, the {Cahn-Hilliard} model is often applied  either to the early stage neglecting the coarsening behavior or vice versa \cite{barton_dynamics_1998}. Also, the spinodal decomposition predicted by the {Cahn-Hilliard} model proceeds until complete segregation of the two phases and no coarsening arrest is possible.

With the {Cahn-Hilliard} model, the effects of mechanical deformation on spinodal decomposition are  neglected, although {J.W. Cahn} already acknowledges the key role of elasticity  in 1961 \cite{cahn_spinodal_1961}. %Recent studies point to the influence of elastic deformations on the coarsening stage and on the characteristic length \cite{style_liquid-liquid_2018, rosowski_elastic_2020, fernandez-rico_elastic_2023}. In particular, it turns out that large differences of elastic properties between the two stable phases tend to reduce the coarsening rate and to turn the channel-like spinodal structure into a pattern where a soft matrix phase encloses stiff droplets \cite{zedalis_precipitation_1986, conley_effect_1989, veron_strain_1997}.
%The coupled Cahn Hilliard elasticity models
A well-known extension of the Cahn-Hilliard model to account for the effect of mechanical deformation is the \textit{Cahn-Larché} model \cite{larche_nonlinear_1978, fratzl_modelling_1999}, where the free energy density reads 

\begin{equation}
    \label{eq:cl_en_generic}
    \psi_{CL}(\bm{\varepsilon},\phi,\nabla\phi) = \psi_{CH}(\phi,\nabla\phi)+\psi_{el}(\bm{\varepsilon}_{el}(\bm{\varepsilon},\phi))\,\quad\text{with}\quad \bm{\varepsilon}_{el}(\bm{\varepsilon},\phi)= \bm{\varepsilon} - \Omega \phi \mathbf{I} \,,
\end{equation}

\noindent where $\bm{I}$ is the identity tensor and $\Omega$ is the chemical expansion coefficient. Here coupling occurs through the volumetric strain associated to the  local phase-field variable. The Cahn-Larché approach is able to qualitatively reproduce the shape of the microstructures arising in metal alloys during coarsening \cite{leo_diffuse_1998, zhu_morphological_2001, garcke_cahn-hilliard_2001, garcke_cahnhilliard_2003, garcke_cahnhilliard_2005, mielke_multiple_2006, jeong_efficient_2015}. Also in this case, however, a single set of parameters cannot predict both early and coarsening stages and their value is heuristically selected so as to reproduce one of the two stages \cite{thompson_spinodal_1997}. In particular, the evolution of the characteristic length scale predicted by the model is not in agreement with the experimental evidence \cite{thompson_spinodal_1997}. Similarly to Cahn-Hilliard, the Cahn-Larché model is not able to predict the arrest of coarsening, although it may lead to a reduction of the coarsening rates depending on the contrast between the elastic parameters of the stable phases \cite{zhu_morphological_2001}. Later in this paper (Sections \ref{CL-1D} and \ref{subsec:2d_standard_formulation}) we also show that the Cahn-Larché model does not predict the scaling of the characteristic length with the stiffness of the dry polymer obtained experimentally in \cite{fernandez-rico_elastic_2023}.

Our minimal model in \cite{fernandez-rico_elastic_2023} is a first attempt to replicate the experimental observations in \cite{fernandez-rico_elastic_2023}. There the Cahn-Hilliard energy is coupled with an extended elastic energy including phase-field dependent elastic properties and an eigenstress-related term at the initial swelling state. Although able to retrieve the correct scaling between initial characteristic length scale and PDMS stiffness, this model lacks a clear justification of the eigenstress-related term and is not able to predict the arrest of coarsening. 

%The Cahn Oono model
The \textit{Cahn-Oono} model, also referred to as the \textit{Ohta-Kawasaki} model, was introduced in \cite{oono_computationally_1987, ohta_equilibrium_1986} to reproduce the arrest of the coarsening stage during phase separation of diblock copolymer mixtures \cite{ohta_equilibrium_1986, bahiana_cell_1990, nishiura_mathematical_1995}. However, its application range was later extended to modeling pattern formation in various systems where nonlocal forces are involved \cite{glotzer_self-consistent_1994, choksi_global_2012, grasselli_cahn-hilliard-oono_nodate}, including the arrest of the Ostwald ripening process in cross-linked polymers \cite{muratov_theory_2002}. In addition to a Cahn-Hilliard-like energetic contribution \eqref{eq:ch_en_generic}, the Cahn-Oono model involves a long-range interaction convolution term competing with the terms inducing coarsening
\begin{equation}
    \label{eq:co_en_generic}
    \psi_{CO}(\phi,\nabla\phi) = \psi_{CH}(\phi,\nabla\phi)+\psi_{conv}(\phi)\,,\quad\text{with}\quad     \psi_{conv}(\phi) = -\frac{\alpha}{2}\int_{\mathcal{B}_0} \phi(\bm{x})\phi(\bm{y})g(\bm{x},\bm{y}) \text{d}\bm{y}\,,
\end{equation}
\noindent where $g(\bm{x},\bm{y})$ is the convolution kernel, $\alpha$ is a parameter determining the magnitude of the long-range interactions and $\mathcal{B}_0$ is the domain occupied by the mixture. A recent paper focusing on modeling of EMPS \cite{qiang_nonlocal_2024} proposes  a 1D Cahn-Oono model with a Gaussian kernel representing  a non-local elastic energy density. Under the assumption that  volumes are conserved during phase separation, an expression for the total strain as a function of the phase variable is obtained, allowing for  analytical results. 
The predictions are compared with the experimental observations in \cite{fernandez-rico_elastic_2023} showing a qualitatively correct correlation between the characteristic length and the polymer stiffness. Clearly, no quantitative comparison is possible with a 1D model. Moreover, the strain energy density based on purely non-local elasticity leads to unphysical non-zero stresses under free deformation  \cite{benvenuti_one-dimensional_2013}.   Also, no conclusions are drawn regarding the evolution of coarsening before arrest.

To the best of our knowledge, beside the Cahn-Larché model and the models in \cite{fernandez-rico_elastic_2023} and \cite{qiang_nonlocal_2024} with their mentioned limitations, no further model is available to predict the coupling between mechanics and spinodal decomposition in general and EMPS in particular. The present study aims to fill this gap and make a step forward towards predictive modeling of EMPS. 

%%%%%% Free energy density Sectop, %%%%%%%%%%%%%%%%%%%%%%%%%%%%%%%%%%%
\section{Free energy density of the proposed model}
\label{app:derivation_coupled_model} 

The dry polymer soaked with HFBMA oil undergoes isotropic swelling into a homogeneous gel which, upon quenching, phase separates with spinodal decomposition (Fig.~\ref{fig:EMPS_process_scheme}). Our goal is to model the evolution of the polymer gel from the isotropically swollen and thermodynamically unstable state right after quenching (i.e., the intermediate configuration) to the current configuration, in which the phase-separated system is characterized by a spinodal microstructure. Modeling of the processes that take place between the reference and the intermediate configurations, including the mixing at high temperature and the quenching, is out of the scope of the present work.

\subsection{Kinematics, chemical variables, free energy density} % -------- Kinemtics / Variables

In the reference configuration, the dry PDMS occupies the volume $\mathcal{B}$ with size $L$, described by the coordinate $\bm{X}$. At $t=t_0$ the system in the intermediate configuration consists of a homogeneous mixture of PDMS and HFBMA oil in an isotropically swollen state (S-PDMS) (Fig.~\ref{fig:EMPS_process_scheme}) occupying the volume $\mathcal{B}_0$ with size $l_0$. The corresponding stretch, deformation gradient and Jacobian are %\footnote{We are implicitly ignoring any consideration on the mixing process between dry polymer and solvent, and treating the transition between reference and intermediate configurations as if it were a pure deformation process of the dry polymer.}
\begin{equation}
\label{eq:def_grad_0}
\delta_0 = l_0/L \quad \bm{F}_0 = \delta_0 \bm{I}\quad\text{and}\quad J_0=\delta_0^3\,.
\end{equation}
\noindent We denote the displacement with respect to the reference state as $\bm{u}_0(\bm{X})=\bm{u}(\bm{X},t_0)$. The deformations due to this initial swelling between reference and intermediate configurations are large (up to 40\% volume increase). The continuum in the intermediate configuration is described by the coordinate $\bm{x}$.
Right after quenching, this intermediate configuration is thermodynamically unstable. The stage at which phase separation has occurred in the swollen gel leading to a spinodal microstructure is considered as the current configuration occupying a domain $\mathcal{B}_c$. Due to the limited variation of concentration taking place during spinodal decomposition, the incremental deformation between the intermediate and the current configuration is assumed to be small, so that linearized kinematics can be adopted, i.e. $\mathcal{B}_c \approx \mathcal{B}_0$ described again by the coordinate $\bm{x}$. 

The displacement field in the current configuration with respect to the reference configuration at any time $t$ is indicated with $\bm{u}(\bm{X},t) = \bm{x}(\bm{X},t) - \bm{X}$. The corresponding deformation gradient, Jacobian and right {Cauchy-Green} deformation tensor are respectively  $\bm{F}(\bm{X},t) = \bm{I} + \nabla_{\!\bm{X}}\bm{u}$, $J = \mathrm{det}(\bm{F})$ and $\bm{C}= \bm{F}^T\bm{F}$. The incremental change of the displacement field between the intermediate and the current states upon phase separation is $\tildbm{u} = \bm{u} - \bm{u}_0$.  In the following we exploit the assumption of small deformations between intermediate and current configurations, i.e $||\nabla_{\!x}\bm{\tilde u}(\bm{x},t)|| \ll 1$, and we introduce the linearized strain tensor associated to $\nabla_{\!x}\tildbm{u}$ 
\begin{equation}
\label{eq:tilde_eps}
        \tildbm{\varepsilon}(\bm{x},t) = \frac{1}{2}\left(\nabla_{\!x}\tildbm{u} + \nabla^T_{\!x}\tildbm{u}\right)\,,
\end{equation}
\noindent along with the linearized skew-symmetric rotation tensor
\begin{equation}
\label{eq:tilde_omega}
        \tildbm{\omega}(\bm{x},t) = \frac{1}{2}\left(\nabla_{\!x}\tildbm{u} - \nabla^T_{\!x}\tildbm{u}\right)\,.
\end{equation}

The oil nominal concentration (i.e. number of oil moles per unit volume of the dry polymer) is denoted as $\Phi(\bm{X},t)$, and its change between the intermediate and the current states reads $\tilde{\Phi} = \Phi - \Phi_0$, where $\Phi_0$ is the homogeneous nominal concentration in the intermediate configuration.
 The number of oil moles per unit current volume of the gel is denoted as $c(\bm{x},t)$ and its change between the intermediate and the current states is $\tilde{c} = c -c_0$, with $c_0$ as the corresponding homogeneous concentration in the intermediate configuration. The relation among the two is $\tilde{\Phi} = J_0 \tilde{c}$. We further introduce the local volume fraction of the solvent $\varphi\in [0,1]$, i.e. the volume of oil per unit volume of gel, with initial homogeneous value $\varphi_0$ in the intermediate configuration. We then rescale it by introducing $\hat \phi=2 \varphi - 1 \in [-1,1]$, with initial value $\hat \phi_0=2 \varphi_0 - 1$ in the intermediate configuration, and finally define the local phase-field variable as
\begin{equation}
    \label{eq:phase-field-variable}
    \phi(\bm{x},t) = 2 \varphi(\bm{x},t) - 1 - \left(2 \varphi_0 - 1\right)  = 2 \left(\varphi(\bm{x},t) - \varphi_0 \right)\,.
\end{equation}
with initial value $\phi_0=0$. The incremental solvent volume fraction $\tilde \varphi = \varphi - \varphi_0$ is directly related to  $\tilde c$ and the solvent molar volume $V_m$ through $\tilde \varphi = V_m\tilde c$.

A central task in developing the model is the choice of a suitable expression for the free energy density of the system. We denote with $\Psi$ the free energy per unit reference volume and with $\psi$ the free energy per unit current volume. The two are related by $\Psi = J_0 \psi$.

\subsection{Elastic energy density and its expansion around the swollen state} %--- Derivation of the elastic energy from expansion
\label{app:linear_elast_MB}
We start by defining the elastic energy density as \cite{bouklas_nonlinear_2015}

\begin{equation}
    \label{eq:elastic_flory_expanded_MB}
    \Psi_{el}(\bm{F}, \Phi) = \underbrace{\frac{1}{2}Nk_BT\left[ \mathrm{tr}(\bm{C}) - 3 -2 \log(J) \right]}_{\Psi_{ent}} + \underbrace{\frac{K}{2}\left[J- (1+\Omega \Phi) \right]^2}_{\Psi_{bulk}}\,,
\end{equation}

\noindent where $N$ is the number of polymer chains per unit reference volume, $k_B$ is the \textit{Boltzmann} constant, $T$ is the temperature, $\Omega$ is the expansion coefficient  and $K$ is the bulk modulus of the material. In \eqref{eq:elastic_flory_expanded_MB}, the first contribution $\Psi_{ent}$ represents the entropic elastic strain energy density, consistent with the Flory-Rehner theory and also used in \cite{hong_theory_2008,bouklas_swelling_2012}, while the second term $\Psi_{bulk}$ governs the compressibility behavior controlled by the bulk modulus $K$ and couples $\Phi$ and $\bm{F}$. If $K\gg Nk_BT$, $\Psi_{bulk}$ can be interpreted as a penalty term that approximately enforces the incompressibility constraint, namely $J \simeq 1+\Omega\Phi$\footnote{In the ideal case of perfectly immiscible solvent and dry polymer, it would be $\Omega=V_m$. However, in reality $\Omega$ is different from $V_m$ and is kept as a constant to be calibrated with experimental results.}. From 
\eqref{eq:elastic_flory_expanded_MB}, the \textit{Cauchy} stress tensor $\bm{\sigma}$ is obtained as
\begin{equation}
    \label{eq:cauchy_stress_flory}
    \bm{\sigma}  = \frac{1}{J} \frac{\partial \Psi}{\partial \bm{F}} \bm{F}^T = \frac{1}{J}Nk_BT\left(\bm{F}\bm{F}^T - \bm{I} \right) + K\left[J- (1+\Omega \Phi) \right]\bm{I}\,
\end{equation}
\noindent 

\noindent and in the intermediate configuration we have
\begin{equation}
    \label{eq:cauchy_stress_flory_int}
    \bm{\sigma}_0   = \frac{1}{J_0}Nk_BT\left(\bm{F}_0\bm{F}_0^T - \bm{I} \right) + K\left[J_0- (1+\Omega \Phi_0) \right]\bm{I}=\bm{0}\quad \Rightarrow \quad J_0-(1+\Omega\Phi_0)=-\frac{Nk_B T}{K J_0}(\delta_0^2-1)\,,
\end{equation}
\noindent where we used \eqref{eq:def_grad_0} and the stress-free assumption at $t=t_0$.

%As in \cite{bouklas_swelling_2012}, 
We now expand \eqref{eq:elastic_flory_expanded_MB} around the homogeneous swollen state up to second order. The detailed steps are reported in \ref{app:linear_elast}. We obtain the expanded energy density
\begin{equation}
    \label{eq:bulk_exp3}
    \begin{split}
\Psi_{el}(\bm{F}, \Phi) &\simeq \Psi_{exp}\left(\tildbm{\varepsilon}, \tilde c\right) =\underbrace{\frac{1}{2}Nk_BT\left[3\delta_0^2 -3 -2 \log(J_0) \right]+\frac{K}{2}\left[J_0-\left(1+\Omega\Phi_0\right)\right]^2}_{\Psi_{0}} +\frac{K}{2}\Omega^2J_0^2\tilde c^2+\\
&-K\Omega J_0 \tilde c\left[J_0-\left(1+\Omega\Phi_0\right)\right] -{KJ_0^2}\Omega\tilde c\,\tr\left(\tildbm{\varepsilon}\right)+\frac{1}{2}\left[{KJ_0^2}-Nk_BT\left(\delta_0^2-1\right)\right]\tr^2\left(\tildbm{\varepsilon}\right)+Nk_BT\delta_0^2\,\tr\left(\tildbm{\varepsilon}^2\right)\,,
\end{split}
\end{equation}
\noindent where $\Psi_{0}$ is the elastic energy density in the intermediate configuration. The expanded elastic energy per unit current volume reads
\begin{equation}
    \label{eq:bulk_exp_inter}
    \begin{split}
\psi_{exp}\left(\tildbm{\varepsilon}, \tilde c\right) =\frac{\Psi_{exp}\left(\tildbm{\varepsilon}, \tilde c\right)}{J_0} =  \underbrace{\frac{\Psi_{0}}{J_0}}_{\psi_{0}} &-K\Omega \tilde c\left[J_0-\left(1+\Omega\Phi_0\right)\right]+\frac{K}{2}\Omega^2J_0\tilde c^2 +\\
&-{KJ_0}\Omega\tilde c\,\tr\left(\tildbm{\varepsilon}\right)+\frac{1}{2}\underbrace{\left[{KJ_0}-Nk_BT\frac{\left(\delta_0^2-1\right)}{J_0}\right]}_{\lambda_0}\tr^2\left(\tildbm{\varepsilon}\right)+\underbrace{\frac{Nk_BT}{J_0}\delta_0^2\vphantom{\left[{KJ_0}-Nk_BT\frac{\left(\delta_0^2-1\right)}{J_0}\right]}}_{G_0}\,\tr\left(\tildbm{\varepsilon}^2\right)\,.
\end{split}
\end{equation}
\noindent The constant parameters $\lambda_0$ and $G_0$ defined in \eqref{eq:bulk_exp_inter} are equivalent Lamé coefficients  defined in analogy with linear elasticity theory. Introducing the phase-field variable, we finally obtain
\begin{equation}
    \label{eq:bulk_exp_pf}
\begin{split}
\psi_{exp}\left(\tildbm{\varepsilon}, \phi\right) &= \underbrace{\frac{1}{2}\lambda_0 \tr^2\left(\tildbm{\varepsilon}\right)+G_0\,\tr\left(\tildbm{\varepsilon}^2\right)\vphantom{\frac{K\Omega}{2V_m}}}_{\hat{\psi}_{el}} -\underbrace{\frac{K J_0 \Omega}{2V_m}}_{m}\phi\,\tr\left(\tildbm{\varepsilon}\right)\underbrace{-\frac{K\Omega}{2V_m} \left[J_0-\left(1+\Omega\Phi_0\right)\right]}_{\tilde{a}_1}\phi+\underbrace{\frac{KJ_0\Omega^2}{8V_m^2}}_{\tilde{a}_2}\phi^2 +\psi_{0}=\\
&=\hat\psi_{el}\left(\tildbm{\varepsilon}\right) -m\phi\,\tr\left(\tildbm{\varepsilon}\right) +\tilde{a}_1\phi+\tilde{a}_2\phi^2  + \psi_{0}\,,
\end{split}
\end{equation}

\noindent where $\hat\psi_{el}\left(\tildbm{\varepsilon}\right)$ is the strain energy density of isotropic linear elasticity evaluated with respect to the intermediate configuration and for  $||\bm{\tilde \varepsilon}(\bm{x},t)|| \ll 1$.

The result we have obtained deserves a few comments. As noted in \cite{bouklas_nonlinear_2015} and mentioned earlier, the relaxation of the incompressibility constraint implied by the second term in \eqref{eq:elastic_flory_expanded_MB} leads to a chemomechanical coupling. Upon expansion of the energy density in \eqref{eq:elastic_flory_expanded_MB} around the swollen state, we obtain the energy density in \eqref{eq:bulk_exp_pf}, which contains the classical strain energy density of linear elasticity for incremental deformations from the swollen state, a coupling term whose strength is governed by parameter $m\ge0$, two chemical terms varying linearly and quadratically with $\phi$, and a constant term (which plays no role and from now on is ignored). Importantly, the second term establishes the coupling between deformation and phase-field variable; it represents the work related to the osmotic pressure, i.e. the pressure induced by the solvent on the polymer mesh within the Biot framework of linear poroelasticity  \cite{bouklas_swelling_2012, biot_general_1941}. Note that the total free energy density in \cite{hong_theory_2008,bouklas_swelling_2012, bouklas_nonlinear_2015}, in addition to the elastic energy, contains a chemical energy; however, the form of this chemical energy is such that the ensuing coupled chemomechanical models in \cite{hong_theory_2008,bouklas_swelling_2012, bouklas_nonlinear_2015} cannot be used to model spinodal decomposition. The same applies to the energy density in \eqref{eq:bulk_exp_pf}, which needs augmentation in order to become able to describe phase separation with spinodal decomposition. This augmentation is performed in the next section.

\subsection{Chemical and interface energy} % -- Adaptation for spinodal decomposition ----------
\label{app:chemical_contr}

The most natural augmentation of the energy density in \eqref{eq:bulk_exp_pf} to endow it with the ability to describe spinodal decomposition is its combination with the Cahn-Hilliard model.  The Cahn-Hilliard free energy density can be written as 
\begin{equation}
\label{CH}
    \psi_{CH}(\phi,\,\nabla \phi) =    \psi_{chem}(\phi) +\psi_{int}(\nabla_{\!x} \phi)\,,
\end{equation}

\noindent where $\psi_{chem}(\phi)$ and $\psi_{int}(\nabla_{\!x} \phi)$ are the chemical (\textit{Ginzburg-Landau}) and interfacial energy densities, respectively, with \cite{provatas_phase-field_2010} 
\begin{equation}
\label{eq:chem_pot}
    \psi_{chem}(\phi) =  \hat{a}_1\phi + \hat{a}_2\phi^2 + \hat{a}_3\phi^3 + \hat{a}_4\phi^4\,,\quad \psi_{int}(\nabla_{\!x} \phi) = \frac{1}{2} \gamma \kappa \left| \nabla_{\!x} \phi \right|^2\,.
\end{equation}

\noindent Here $\hat{a}_i$ are constant model parameters controlling the shape of the Ginzburg-Landau double-well potential, $\kappa$ is a parameter with the dimension of a square length governing the thickness of the regularized interface between the phases (this thickness being proportional to $\sqrt{\kappa}$) and $\gamma$ is a parameter with the dimension of an energy density to be motivated later. Note that in $\psi_{chem}$ the linear term in $\phi$ can be neglected since it gives a constant contribution to the chemical potential and, therefore, does not influence the phase separation process. For the same reason, we can also neglect the linear term in $\phi$ within \eqref{eq:bulk_exp_pf}. Moreover, the cubic term can be dropped in case of a symmetric double-well potential. Due to lack of sufficient information on the shape of the double-well potential in EMPS, we consider the simplest form of the model and ignore the cubic term. We also remark that, in order for \eqref{eq:chem_pot}$_1$ to represent a double-well potential and, hence, lead to phase separation, it must be $\hat{a}_2<0$ and $\hat{a}_4>0$. In the literature, $\hat{a}_2$ is often related to the difference between incubation temperature and room temperature during quenching.

\subsection{Long-range interaction energy} \label{app:1d_convolution_derivation}% -- The Long range term--------------

In the proposed model, a long-range interaction energy is introduced following the Cahn-Oono model as
\begin{equation}
    \begin{split}
    \label{eq:conv_term}
        \psi_{conv}\left(\phi\right)=-\frac{\alpha}{C} \int_{\mathcal{B}_0} \phi(\bm{x}) \phi(\bm{y}) g(\bm{x}, \bm{y}) \text{d}\bm{y}\,,
    \end{split} 
\end{equation}
\noindent where $g(\bm{x}, \bm{y})$ is a kernel function and $C$ is a constant. In 2D and 3D, the kernel is defined as the {Green}’s function for the {Laplace}’s equation $\Delta g(\bm{x},\bm{y}) = \delta(\bm{x},\bm{y})$, while $C=2$. This leads to a linear term $-\alpha\phi$ appearing in the mass balance equation.  To obtain the same effect in the 1D setting, the distance function $\left|x-y\right|$ should be adopted as kernel with $C=4$. In summary, the kernel function is defined as  \cite{ohta_equilibrium_1986}

\begin{equation}
    g(\bm{x},\bm{y}) = 
    \begin{cases}
    \left|x-y\right| & \text{in 1D,}\\
        \displaystyle\frac{1}{2\pi}\mathrm{ln}\left(|\bm{x} - \bm{y}|\right), & \text{in 2D,}  \\
        \displaystyle-\frac{1}{4\pi |\bm{x} - \bm{y}|}, & \text{in 3D,} \, 
    \end{cases} \quad   \text{and}\quad   C = 
    \begin{cases}
   4& \text{in 1D,}\\
        2, & \text{in 2D,}  \\
        2, & \text{in 3D.} \, 
    \end{cases}
\end{equation}

%\indent In higher dimensions, we have
%\begin{equation}
 %   \psi_{conv}(\phi) = -\frac{\alpha}{2}\int_{\mathcal{B}_0} \phi(\bm{x})\phi(\bm{y})g(\bm{x},\bm{y}) \text{d}\bm{y}\,.
%\end{equation}

The kernel in 3D is referred to as {Coulomb}-type kernel, usually adopted to represent repulsive forces \cite{choksi_2d_2011}. In the 1D and 2D cases the forms are no longer of Coulomb type, however, they still represent repulsive forces since they preserve a similar final effect on the mass balance equation. $\psi_{conv}$ is meant to describe finite-distance interactions within the crosslinked mesh, i.e. it accounts for the fact that the deformation at each crosslink depends on the state of the surrounding ones.

\subsection{Summary of the free energy density} % -- Summary of the adopted free energy density
\label{app:final model}

Considering \eqref{eq:bulk_exp_pf}, \eqref{CH}, \eqref{eq:chem_pot} and \eqref{eq:conv_term} 
along with the assumptions in \ref{app:linear_elast_MB} and \ref{app:chemical_contr} 
we obtain the following expression for the total free energy density
    \begin{equation}
        \label{eq:int_energy}
        \psi(\tildbm{\varepsilon},\, \phi,\, \nabla_{\!x} \phi) = \underbrace{\hat\psi_{el}\left(\tildbm{\varepsilon}\right)-m\phi\,\tr\left(\tildbm{\varepsilon}\right)}_{\psi_{el}(\tildbm{\varepsilon},\phi)}+ \underbrace{a_2\phi^2 + a_4\phi^4}_{\psi_{chem}(\phi)}+ \frac{1}{2} \gamma \kappa \left| \nabla_{\!x} \phi \right|^2 + \psi_{conv}\left(\phi\right)\,
    \end{equation}
\noindent with $a_2 = \tilde{a}_2 + \hat{a}_2$ and $a_4=\hat{a}_4$. We then redefine $a_2=\gamma \xi/2$ and $a_4=\gamma \beta$. The parameter $\xi$ encodes the dependence on the quenching temperature. Following \cite{novick-cohen_nonlinear_1984,mukherjee_statistical_2021} and assuming that the \EMPS process is isothermal, we take
    \begin{equation}
        \label{eq:xi_T}
        \xi = \frac{T_{R} - T_{inc}}{T_{inc}}\,.
    \end{equation}
The parameter $\gamma>0$, with dimensions of an energy density, determines the relative importance of $\psi_{chem}$ compared to $\psi_{el}$. As better detailed in Section \ref{app:1d_calibration}, for large values of $\gamma$ the influence of the mechanical part on the spinodal decomposition vanishes, while for small values of $\gamma$ it is the mechanical contribution that mainly governs the characteristics of the  spinodal structures.  The dimensionless parameter $\beta$ controls the so-called \textit{binodal points}, namely the values of $\phi=(\phi_b^1,\,\phi^2_b)$ composing the spinodal structures at thermodynamic equilibrium.  

Finally, we obtain the following form for the total free energy density of the proposed model:
    \begin{equation}
        \label{eq:int_energy_final}
        \psi(\tildbm{\varepsilon},\, \phi,\, \nabla_{\!x} \phi) = \underbrace{\hat\psi_{el}\left(\tildbm{\varepsilon}\right)-m\phi\,\tr\left(\tildbm{\varepsilon}\right)\vphantom{\frac{1}{2}\gamma\kappa \left| \nabla_{\!x} \phi \right|^2}}_{\psi_{el}\left(\tildbm{\varepsilon},\phi\right)}+  \underbrace{\frac{1}{2}\gamma\left(\xi\phi^2 + 2\beta\phi^4\right)}_{\psi_{chem}\left(\phi\right)}+  \underbrace{\frac{1}{2}\gamma\kappa \left| \nabla_{\!x} \phi \right|^2}_{\psi_{int}\left(\nabla_{\!x} \phi\right)}+ \psi_{conv}\left(\phi\right)\,.
    \end{equation}
Thus, the proposed free energy density includes a poroelastic energy density obtained by expanding the model proposed in \cite{bouklas_nonlinear_2015} around the intermediate swollen configuration, a Cahn-Hilliard chemical energy density governing spinodal decomposition and a Cahn-Oono long-range interaction energy density.
%%%%%% 1D analysis of EMPS %%%%%%%%%%%%%%%%%%%%%%%%%%%%%%%%%%%%%
\section{1D model of EMPS}  
\label{sec:1d_model} 

In this section, starting from the free energy density  \eqref{eq:int_energy_final}, we formulate and solve the boundary value problem of EMPS in 1D.

\subsection{Problem formulation}  % ---------- 1D Problem formulation  ------------ 
\label{subsec:1d_formulation} 

We consider here a 1D version of the system schematized in Fig.~\ref{fig:EMPS_process_scheme}. In the reference configuration, the dry polymer is represented by a bar occupying the domain $\mathcal{B}=[0, L]$ described through the coordinate $X$.  As mentioned in Section \ref{sec:Introduction}, a gel is then obtained by soaking the dry polymer in a solvent ({S-PDMS}), after which the system state is homogeneous in space and characterized by a swollen length $l_0>L$. This state corresponds to the intermediate configuration $\mathcal{B}_0=[0, l_0]$, described by the current coordinate $x$.
As initial state at the time instant $t_0=0$ we consider the S-PDMS right after quenching, i.e rapid cooling to room temperature $T_{R}=20 \degree C$. Then, the gel undergoes spinodal decomposition. This stage corresponds to the current configuration, which is associated with the current coordinate $x$ in the domain $\mathcal{B}_c=[0, l(t)]$ and evolves within a time interval $t\in[0,T]$. We further assume that the phase separation process takes place under isothermal conditions at $T_{R}$. The incremental displacement field between the intermediate and the current configuration is denoted as $\tilde{u}(x,t)$, and the associated infinitesimal incremental strain field is $\tilde{\varepsilon} = d\tilde{u}/dx$. 

As follows, we introduce the adopted free energy density and the resulting constitutive equations, and conclude the section with the governing equations for the proposed 1D model. We consider the 1D counterpart of the free energy density in \eqref{eq:int_energy_final}:
\begin{equation}
    \label{eq:1d_tot_energy}
    \psi\left(\phi, \phi_{,x}, \tilde{\varepsilon} \right) =  \psi_{chem}(\phi) + \psi_{int}\left( \phi_{,x} \right) + \psi_{el}(\phi, \tilde{\varepsilon}) + \psi_{conv}(\phi)\,,
\end{equation}
\noindent where 
    \begin{equation}
        \label{eq:1d_bulk_energy}
        \psi_{el} = \frac{1}{2} E_0 \tilde{\varepsilon}^2 - m \phi \tilde{\varepsilon},\quad\psi_{chem} = \gamma \left( \frac{1}{2} \xi \phi^2 + \beta \phi^4 \right),\quad         \psi_{int} = \frac{1}{2}\gamma \kappa \left| \frac{\partial \phi}{\partial x} \right|^2,\quad         \psi_{conv}  = -\frac{\alpha}{4}\phi(x,t)\int_0^{l_0} \phi(y,t) g(x,y) \,\text{d}y.
    \end{equation}
Hence, the total free energy density in the 1D setting has the following form   
\begin{equation}
    \label{eq:1d_tot_energy_developped}
    \begin{split}
        \psi\left(\phi, \phi_{,x}, \tilde{\varepsilon} \right)  =  \gamma \left( \frac{1}{2} \xi \phi^2 + \beta \phi^4 \right) + \frac{1}{2}\gamma \kappa &\left| \phi_{,x} \right|^2 
        + \frac{1}{2} E_0 \tilde{\varepsilon}^2 -m\phi \tilde{\varepsilon} -\frac{\alpha}{4}\phi(x,t)\int_0^{l_0} \phi(y,t) |x-y|\,\text{d}y\,.
    \end{split}
\end{equation}

The constitutive equations are now derived from  \eqref{eq:1d_tot_energy_developped}. In particular, the conjugates of the phase-field variable and the strain tensor are the chemical potential $\tilde \mu$ and the Cauchy stress $\tilde \sigma$, which read
\begin{equation}
    \hspace{0 mm}
    \label{eq:1d_chemical_potential_cpld}
    \begin{split}
        \tilde \mu & =\frac{\partial \psi_{chem}}{\partial \phi} - \frac{\partial}{\partial x} \left( \frac{\partial \psi_{int}}{\partial \phi_{,x}} \right) + \frac{\partial \psi_{el}}{\partial \phi} + \frac{\partial \psi_{conv} }{\partial \phi}=  \gamma \left(\xi \phi + 4\beta\phi^3 \right) - \gamma \kappa \phi_{,xx} -m\tilde{\varepsilon} - \frac{\alpha}{2} \int_0^{l_0} \phi(y,t) \left| x-y \right| \, \text{d}y \,,
    \end{split}
\end{equation}
and
\begin{equation}
    \hspace{0 mm}
    \label{eq:1d_stress_strain}
    \tilde \sigma  = \frac{\partial \psi_{el}}{\partial \tilde{\varepsilon}}
    =   \underbrace{E_0 \tilde{\varepsilon}   \vphantom{-m\phi}}_{\tilde \sigma_{el}} \, - \, \underbrace{m\phi}_{\tilde\sigma_{os}}\,,
\end{equation}
\noindent respectively. The stress \eqref{eq:1d_stress_strain} includes the elastic stress $\tilde \sigma_{el}$, and the osmotic pressure $\tilde \sigma_{os}$ \cite{bouklas_swelling_2012}. 
Starting from \eqref{eq:1d_chemical_potential_cpld}, we then define the flux $J$ as
\begin{equation}
    \label{eq:1d_flux}
    J = -M \frac{\partial\tilde \mu}{\partial x}= -M \left[ \gamma \left( \xi +12\beta \phi^2 \right)\phi_{,x} - \gamma \kappa \phi_{,xxx} - m\tilde{\varepsilon}_{,x}-\frac{\alpha}{2}\left( \int_0^x\phi(y,t)\text{d}y - \int_x^{l_0} \phi(y,t)\text{d}y \right) \right],
\end{equation}
\noindent where $M$ is the mobility coefficient that is here assumed constant. 
We can then write the mass balance equation, which yields the following fourth-order differential equation
\begin{equation}
    \label{eq:1d_mass_balance_phi}
    \frac{\partial \phi}{\partial t}=-\frac{\partial J}{\partial x} = M \left[\gamma \left( \xi + 12\beta\phi^2 \right) \phi_{,xx} + 24\gamma\beta \phi \phi_{,x}^2 -\gamma \kappa \phi_{,xxxx}
    -m\tilde{\varepsilon}_{,xx}  - \alpha \phi \right]\,,
\end{equation}

\noindent containing the coupling term $-m\varepsilon_{,xx}$. The term $-\alpha \phi$ is typical of the so-called \textit{Swift-Hohenberg} model responsible for coarsening arrest \cite{politi_dynamics_2014}. We remark that the main difference between the present and the Swift-Hohenberg approach lies in the form of the other nonlinear terms. 

For phase transition problems in solid materials it is often assumed that the time scale associated with the migration of the species is much larger than the time scale to achieve mechanical equilibrium, so that the latter can be considered instantaneous \cite{garcke_cahn-hilliard_2001}. Thus, for the mechanical problem we assume quasi-static conditions. The first-order optimality condition  of \eqref{eq:1d_tot_energy_developped} with respect to $\tilde u$ leads to the mechanical equilibrium equation
\begin{equation}
    \label{eq:1d_linear_momentum}
    \frac{\partial \tilde \sigma}{\partial x} = 0\,, \quad \forall \ t\in[0,T]\,,
\end{equation}
\noindent For the mechanical  boundary conditions, we assume a bar $\mathcal{B}_0$ clamped at the left end and free at the right end. Hence, we have
\begin{equation}
    \label{eq:mech_BC}
    \tilde{u}(0,t)=0\,, \quad \tilde \sigma(l_0,t) = 0 \,,\quad \forall t \in [0,T]\,.
\end{equation}
We assume the following boundary conditions for the chemical part
\begin{equation}
    \label{eq:1d_zero_flux}
    J(0,t) = J(l_0,t) = 0\,, \quad \phi_{,x}(0,t) = \phi_{,x}(l_0,t) = 0 \,, \quad \forall t \in [0,T]\,,
\end{equation}
\noindent while the initial condition is 
\begin{equation}
    \label{eq:1d_initial_condition}
    \phi(x,0) = \phi_0 = 0 \,,\quad\forall x \in \mathcal{B}_0\,.
\end{equation}

 Substituting \eqref{eq:1d_linear_momentum} in \eqref{eq:mech_BC} gives
\begin{equation}
    \label{eq:0stress}
    \tilde \sigma(x,t) = 0\,, \quad \forall (x,t)\in \mathcal{B}_0\times [0,T]\,,
\end{equation}
\noindent which can be further combined with \eqref{eq:1d_stress_strain} to write the total strain as a function of the phase-field variable
\begin{equation}
    \label{eq:1d_strain}
    \tilde{\varepsilon} = \frac{m}{E_0}\phi =\tilde \varepsilon_{os} \,,
\end{equation}

\noindent where $\tilde \varepsilon_{os}$ is the osmotic contribution. The latter can be rewritten as
\begin{equation}
    \label{eq:1d_volumetric_strain}
    \tilde{\varepsilon}_{os}= \frac{m}{E_0}\phi  = \frac{2m}{E_0}\tilde{\varphi}  = \Omega\tilde{\varphi}\,,
\end{equation}
\noindent where $\Omega=2m/E_0$ is the volumetric expansion coefficient in the intermediate configuration. As detailed in \ref{app:swelling_coefficient}, this can be estimated using the experimental measurements in \cite{fernandez-rico_elastic_2023}. 

\subsection{Comparison with Cahn-Larché model} % ------------ Cahn Larché 1D ------
\label{CL-1D}
Within the 1D setting, the Cahn-Larché model reads
\begin{equation}
\label{eq:1d_standard_model_free_energy}
    \psi_{CL}( \mathbf{\tilde \varepsilon}, \phi, \nabla_{\!x} \phi) = \gamma \left( \frac{1}{2}\xi\phi^2 + \beta \phi ^4 \right) + \frac{1}{2}\gamma \kappa |\nabla_{\!x} \phi|^2 + \frac{1}{2} E_0  \left(\tilde\varepsilon - \Omega \phi \right)^2\,.
\end{equation}
The stress-free condition \eqref{eq:0stress} leads to $\tilde{\varepsilon}(x,t) =  \Omega\phi(x,t)$. This makes the elastic energy contribution vanish regardless of whether the Young's modulus depends on the phase-field  variable or not. Hence, the Cahn-Larché model simplifies to the (uncoupled) Cahn-Hilliard model, where the elastic deformations have no effect on spinodal decomposition. Note, however, that this is no longer true for general mechanical boundary conditions (different from \eqref{eq:mech_BC}) nor in 2D (see Section \ref{subsec:2d_standard_formulation}).

\subsection{Early-stage characterization}  % ------ Early stage characterization -------------
\label{subsec:1d_linear_stability} 

At the early stage of spinodal decomposition, the initial  homogeneous mixture is thermodynamically unstable and  phase separates, giving rise to a spinodal microstructure. Its emergence occurs at an initial characteristic time $\tau$ and with an initial characteristic length $\ell_0$. In the 1D setting, these parameters can be analytically obtained by performing a linear stability analysis of the mass balance equation. To this end, we consider an initial mixture with $\phi(x,0)=\phi_0=0$ and we apply the perturbation
\begin{equation}
    \label{eq:1d_perturb}
    \delta \phi(x,t)=e^{\omega t + i k x}\,,
\end{equation}
where $\omega$ and $k$ are the pulsation and the wave vector, respectively. After the linearization of \eqref{eq:1d_mass_balance_phi} around $\phi_0$ and the substitution of \eqref{eq:1d_perturb}, we obtain
\begin{equation}
    \label{eq:1d_linear_perturbation}
    \frac{\partial }{\partial t}\left[\phi_0+\delta \phi\left(x,t\right)\right]=\frac{\partial }{\partial t}\delta \phi(x,t)= {M} \left( h_0 \delta\phi_{,xx}  - \gamma \kappa \delta\phi_{,xxxx} - \alpha \delta \phi  \right)\,,
\end{equation}
\noindent where
\begin{equation}
    \label{eq:1d_h_0}
    h_0 = \gamma\xi - \frac{m^2}{E_0}\,.
\end{equation}
From  \eqref{eq:1d_linear_perturbation}, we retrieve the wave dispersion equation
\begin{equation}
    \label{eq:1d_dispersion_equation}
    \omega = -{M} \left(  h_0 k^2 + \gamma \kappa k^4 + \alpha   \right)\,.
\end{equation}
\noindent Following \eqref{eq:1d_dispersion_equation}, the case $\omega>0$ means that the initial homogeneous condition $\phi(x,0)=\phi_0=0$ is unstable and leads to  phase separation. Conversely, if $\omega<0$ the perturbation \eqref{eq:1d_perturb} is damped and no spinodal decomposition occurs. Since $\gamma\ge0$, $\kappa\ge0$ and $\alpha\ge0$, the sign of $\omega$ is controlled by $h_0$ that, following \eqref{eq:1d_h_0}, is always non-positive since $\xi\le0$.

As follows, we separate the analysis of \eqref{eq:1d_dispersion_equation} distinguishing the cases where $\alpha=0 $ and $\alpha>0$. In the case $\alpha=0$ the long-range interactions are neglected and the chemical energy terms are those of the classic Cahn-Hilliard equation. On the other hand, for $\alpha>0$ the long-range interactions are accounted for and the chemical behavior is influenced by the Cahn-Oono term. We denote the case $\alpha=0$ as \textit{Cahn-Hilliard model coupled with elasticity}, and the case $\alpha>0$ as \textit{Cahn-Oono model coupled with elasticity}.

\subsubsection{Case ${\alpha = 0}$ (Cahn-Hilliard model coupled with elasticity).} %---- Cahn Hilliard elasticity

\begin{figure}[h!]
    % \hspace*{-2cm}
    \centering
    \captionsetup{justification=centering}
    \includegraphics[scale = 0.8]{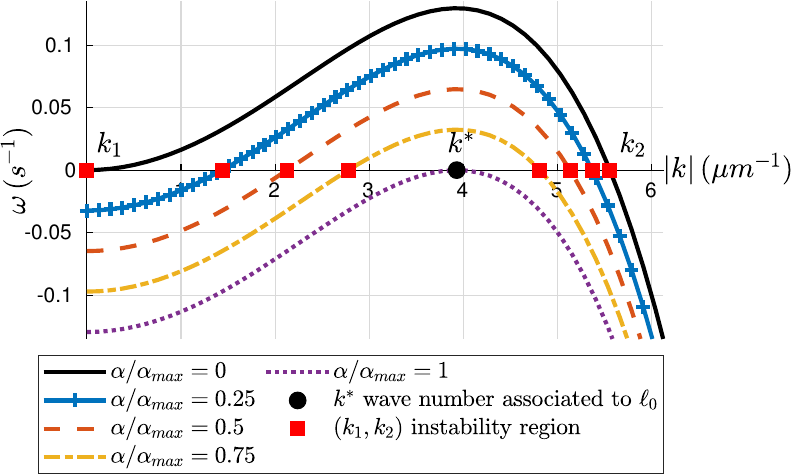}
    \caption{Dispersion relation  \eqref{eq:1d_dispersion_equation} for $E_{PDMS}=350 \, kPa$ and for different values of $\alpha$.}
    \label{fig:1d_pulsation_k_cahn_hilliard}
\end{figure}

As illustrated in Fig.~\ref{fig:1d_pulsation_k_cahn_hilliard}, which gives $\omega$ as a function of $|k|$, in this case $\omega > 0 $ for  $|k|\in(0, k_2)$ with 
\begin{equation}
    \label{eq:1d_k1_k2}
    k_2 = \left|\sqrt{ \frac{m^2}{\gamma \kappa E_0}-\frac{\xi}{\kappa}}\right|\,.
\end{equation}
Within the interval $|k|\in(0, k_2)$, the perturbations are amplified and trigger  spinodal decomposition. The early-stage characteristic length of the spinodal structure $\ell_0$ can be approximated as the wavelength associated to the wave number $k^*$ which maximizes \eqref{eq:1d_dispersion_equation} (Fig.\ref{fig:1d_pulsation_k_cahn_hilliard}) and reads
\begin{equation}
    \label{eq:1d_initial_characteristic_length}
    \ell_0 = \frac{2\pi}{k^*} = 2\pi \sqrt{-\frac{2\gamma \kappa}{h_0}} = 2\pi \sqrt{\frac{-2\kappa}{\xi -m^2/(\gamma E_0 )  }\,.}
\end{equation}

\noindent Also, the initial characteristic time $\tau$ after which the spinodal structure is expected to emerge can be evaluated as
\begin{equation}
    \label{eq:1d_initial_characteristic_time}
    \begin{split}
        \tau = \frac{2\pi}{\omega (k^*)} = \frac{8\pi\gamma \kappa}{M h_0^2}  =  \frac{8\pi\kappa}{M \gamma \left[\xi -m^2 /(\gamma E_0)\right]^2 }\,.
    \end{split}
\end{equation}

\subsubsection{Case ${\alpha > 0}$ (Cahn-Oono model coupled with elasticity).} % ------- Cahn Oono elasticity ---------

The term $\alpha>0$ in  \eqref{eq:1d_dispersion_equation} reduces the value of $\omega$ by shifting the pulsation curve downwards of an amount proportional to the value of $\alpha$ (Fig.~\ref{fig:1d_pulsation_k_cahn_hilliard}). Hence, the range of unstable wave numbers reduces to $|k|\in(k_1, k_2)$ with 
\begin{equation}
    \label{eq:1d_k1_k2_alpha}
    k_1 =\left| \sqrt{ \frac{-h_0-\sqrt{h_0^2-4\gamma\alpha\kappa}}{2\gamma\kappa}}\right|\,, \quad k_2 = \left|\sqrt{ \frac{-h_0+\sqrt{h_0^2-4\gamma\alpha\kappa}}{2\gamma\kappa}}\right|\,,
\end{equation}
There exists a limit value
\begin{equation}
    \label{eq:1d_alpha_max}
    \alpha_{max} = \frac{1}{4\gamma\kappa}  \left( \gamma\xi -\frac{m^2}{E_0}\right)^2 
\end{equation}
beyond which $\omega$ is negative for any wave number and the homogeneous mixture remains thermodynamically stable. 

\noindent Since the pulsation curve is only translated along the $\omega$ axis  (Fig.~\ref{fig:1d_pulsation_k_cahn_hilliard}), the values of $k^*$ and of the initial characteristic length $\ell_0$ in \eqref{eq:1d_initial_characteristic_length} do not change. Conversely, the initial characteristic time increases and reads
\begin{equation}
    \label{eq:1d_initial_characteristic_time_oono}
    \tau \sim \frac{2\pi}{\omega (k^*)} = \frac{8\pi\gamma\kappa}{M \left( h_0^2 -4\gamma\kappa\alpha \right)}
\end{equation}

\subsection{Numerical results}    % -------- 1D Numerical Results -----------
\label{subsec:1d_numerical_results}

In this subsection, we compare the predictions of the proposed model with the main experimental observations summarized in Section \ref{sec:Introduction}. Clearly, this is only a qualitative comparison due to the 1D setting. The approximate solution of the system of governing equations \eqref{eq:1d_mass_balance_phi}-\eqref{eq:1d_linear_momentum} is obtained using the finite element method, where the domain $\mathcal{B}_0$ is a  bar with length 30 $\mu m$ discretized with 300 linear truss elements. The other computational aspects relevant for the following analyses are summarized in \ref{app:solution_strategy}.

\subsubsection{Parameter calibration}  % ------- Experimental and numerical parameters -------------
\label{subsubsec:1d_calibration}

Tab.~\ref{tab:1d_table_num_data} summarizes the parameters used in the numerical computations, calibrated from the experimental data in Tab.~\ref{tab:table_PDMS_exp_data}. The detailed description of the calibration procedure is reported in \ref{app:1d_calibration}. 
The chemical parameters $\gamma, \,  \kappa$ are calibrated for the \PDMS with $E_{PDMS}=350 \, kPa$ and then extended to the other materials to assess the predictive capability of the model. They are obtained through qualitative considerations and by matching the value of the initial characteristic length given by \eqref{eq:1d_initial_characteristic_length} with the experimental observation. We recall that \cite{fernandez-rico_elastic_2023} measured the arrested characteristic length of the spinodal structure, and it is not known whether this coincides with the initial length or the result of some coarsening. For a comparison with the results illustrated in the following, we assume that the experimental length is the initial one. The parameter $\beta$ is obtained for all mixtures employing the Maxwell common tangent method \cite{gibbs_equilibrium_1878} and considering the binodals $(\phi_b^1,\phi_b^2)=(-0.1,0.1)$ (see \ref{app:1d_calibration}). The estimates for the mobility coefficients $M$ are extrapolated from the data reported in \cite{rosowski_elastic_2020}, where a polymeric material similar to the one considered here is studied. The parameter $\xi$ is given by  \eqref{eq:xi_T} with $T_{inc}=60 \degree C$ and $T_{R}=20 \degree C$. The swelling ratio $s_0$ is used to estimate the expansion coefficient using \eqref{eq:1d_volumetric_strain} and the relations in \ref{app:swelling_coefficient}.

\begin{table}[]
    \centering
    \begin{tabular}{ccccccc}
$E_{PDMS}$ & $\gamma$    & $\xi$     & $\beta\,^{(\dagger)}$  & $\kappa$   &  $m$   & $M$  \\
$(kPa)$    & $(Jm^{-3})$ & $(-)$     & $(-)$   & $(m^2)$  & $(kPa)$ & $(m^{3}\, s \, kg^{-1})$ \\
$180$      & $2.5 \, 10^3$      & $-0.12$   & $16.94$ & $4.85\, 10^{-14}$ & $7.00$  & $5.77 \, 10^{-18}$ \\
$350$      & $2.5 \, 10^3$      & $-0.12$   & $32.53$ & $4.85\, 10^{-14}$ & $14.30$  & $5.17 \, 10^{-18}$ \\
$800$      & $2.5 \, 10^3$      & $-0.12$   & $80.57$ & $4.85\, 10^{-14}$ & $34.92$ & $3.32 \, 10^{-18}$\\
        \multicolumn{7}{l}{\footnotesize$^{(\dagger)}$ The value adopted here is defined by \eqref{eq:1d_beta_star} in \ref{app:1d_calibration}, where it takes the name $\beta^*$.}
    \end{tabular}
    \caption{\label{tab:1d_table_num_data}Numerical parameters for three PDMS stiffnesses.}
\end{table}

\subsubsection{Case $\alpha = 0$ (Cahn-Hilliard model coupled with elasticity)} % ---------- Cahn Hilliard based results  ------
\label{subsubsec:1d_cahn_hilliard_results}

\begin{figure}[!hbt]
    \centering
    \captionsetup{justification=centering}
    \includegraphics[width=\textwidth]{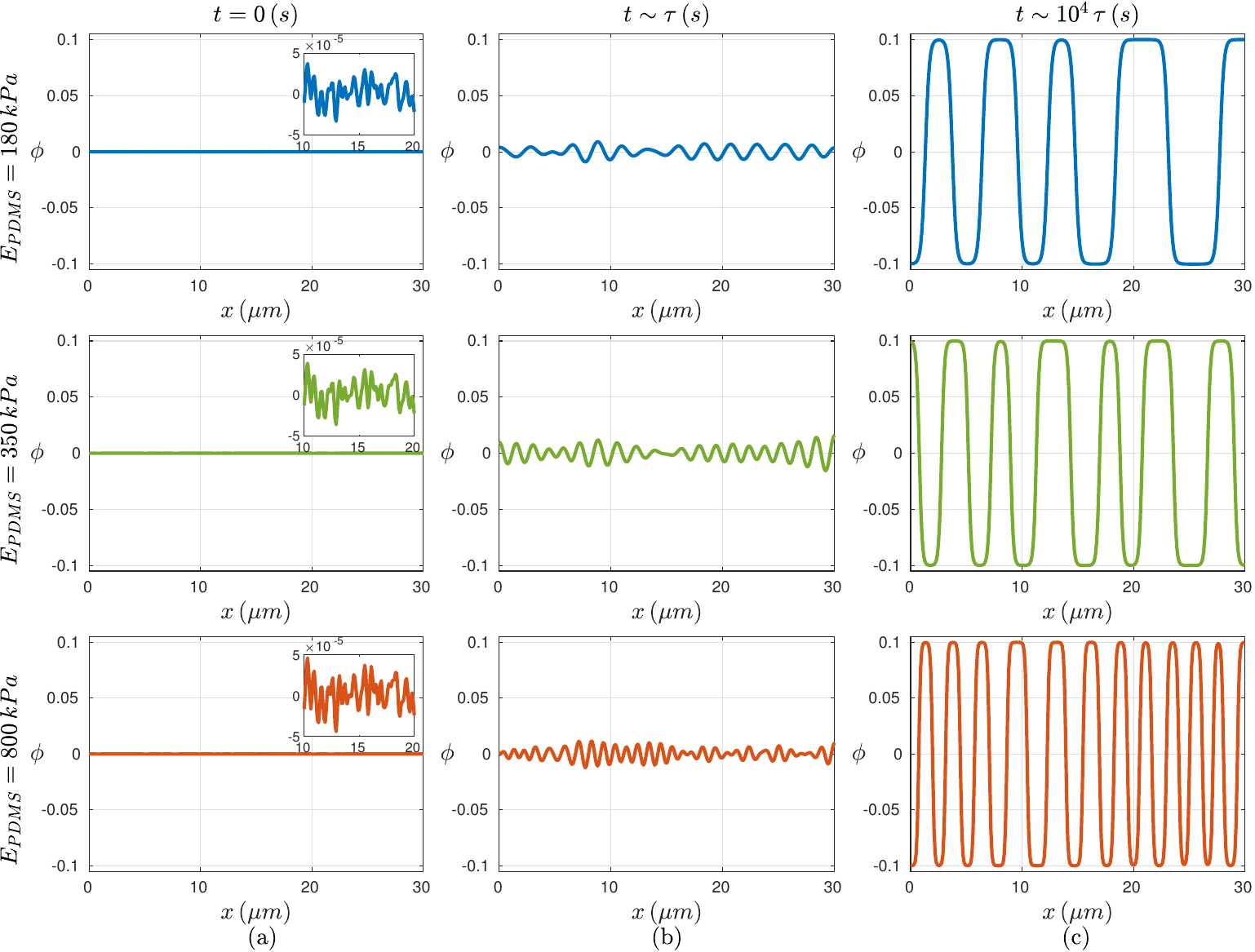}
    \caption{$\phi(x,t)$ profile at $t = 0 \, (s)$ (a), $t \sim \tau \, (s) $ (b) and $t \sim 10^4\tau \, (s)$ (c) for different PDMS stiffnesses (rows) in the  model. Insets display a magnified view of the perturbation at the initial state.}
    \label{fig:1d_PDMS_morphologies}
\end{figure}

We first study spinodal decomposition under the assumption $\alpha=0$. Initially, the three mixtures with $E_{PDMS} = 180, \, 350, \, 800 \, kPa$ share the same perturbed profile $\phi(x,t=0) = 0 + \delta \phi(x)$, where $\delta \phi(x)$ follows a uniform probability distribution in the interval $(-5 \times 10^{-4}, 5\times 10^{-4})$ with $\int_0^{l_0}\delta\phi(x)\,\text{d}x = 0$ (Fig.~\ref{fig:1d_PDMS_morphologies}a). Different evolution stages of the phase-field variable for a given perturbation are illustrated in Fig.~\ref{fig:1d_PDMS_morphologies}. At early stages, i.e. for $t \sim \tau$, the mixtures lose stability and phase separation starts leading to a spinodal structure with initial characteristic length $\ell_0$  (Fig.~\ref{fig:1d_PDMS_morphologies}b). 

\begin{figure}[!hbt]
    \centering
    \captionsetup{justification=centering}
    \includegraphics[scale = 0.7]{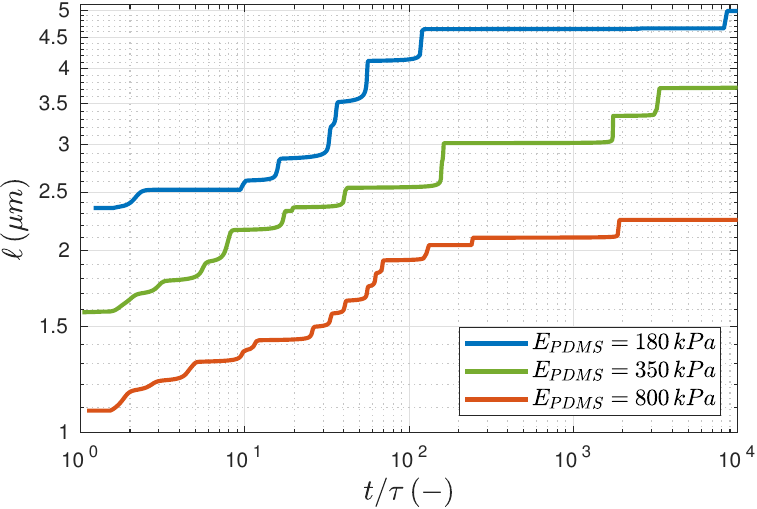}
    \caption{Evolution of the characteristic length for the three different PDMS stiffnesses for a given initial perturbation in the 1D model.}
    \label{fig:1d_ell_t}
\end{figure}

\begin{figure}[!hbt]
    \centering
    \captionsetup{justification=centering}
    \includegraphics[scale = 0.7]{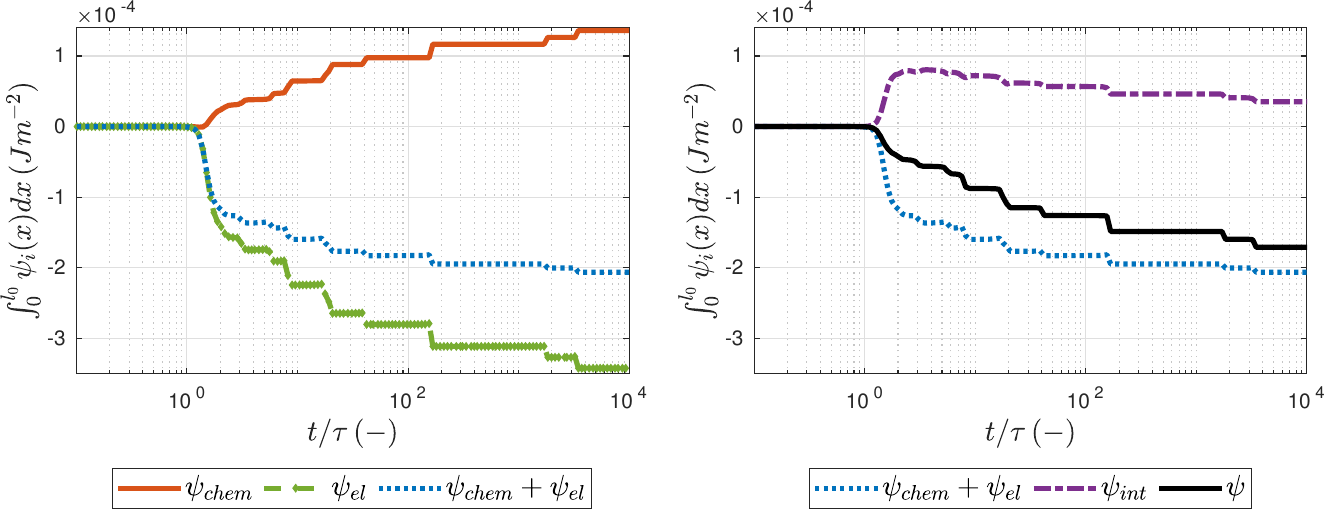}
    \caption{Evolution of the total energy and of its local and nonlocal contributions for the $350 \, kPa$ PDMS mixture for a given initial perturbation in the 1D model with $\alpha=0$.}
    \label{fig:1d_PDMS_energies}
\end{figure}

At the end of the early stage, the domain is composed of clusters separated by (regularized) interfaces where the phase-field variable reaches its binodal equilibrium values. The mixture then enters the coarsening stage where the characteristic length increases over time to form larger domains at the expense of the number of interfaces. An illustrative step of the coarsening stage is shown in Fig.~\ref{fig:1d_PDMS_morphologies}c, while the evolution of the characteristic length evaluated with the fast Fourier transform method in \cite{zhu_coarsening_1999} is reported in  Fig.~\ref{fig:1d_ell_t}. We observe that $\ell$ increases in steps, each representing the disappearance of one or more interfaces due to the coarsening process. Unlike in the experiments  \cite{fernandez-rico_elastic_2023}, coarsening is not arrested even at a late stage of the process. This can be explained considering that the interfaces represent the main  energetic cost at this stage, hence the system evolves reducing their number until reaching thermodynamic equilibrium. This is further demonstrated in Fig.~\ref{fig:1d_PDMS_energies}, which illustrates the evolution of the total energy and its different contributions. The initial sub-horizontal branch pertains to the initial state characterized by an unstable homogeneous mixture. The emergence of the spinodal structure with $\ell \sim \ell_0$ at $t \sim \tau$ occurs with a decrease of energy confirming that phase separation is energetically favorable compared to the homogeneous mixture. At the same time, however, the interfacial energy rises substantially due to the large number of formed interfaces (Fig.~\ref{fig:1d_PDMS_energies}). This leads to the coarsening stage also termed  \textit{Ostwald ripening} \cite{voorhees_ostwald_nodate, chen_dynamics_1993, ostwald_studien_1897}, during which the system evolves toward lower energy states by decreasing the number of interfaces and, with it, the interfacial energy contribution $\int_0^{l_0} \psi_{int}(x) \text{d}x$. Further, the formation of larger stable domains with composition equal to the binodal points has a stabilizing effects since the term $\int_0^{l_0} \left( \psi_{chem}(x)+ \psi_{el}(x) \right) \text{d}x$ decreases as well. If no balancing contribution is introduced, this process progresses until a single interface dividing two large clusters of stable phases remains.

\begin{figure}[h!]
    % \hspace*{-2cm}
    \centering
    \captionsetup{justification=centering}
    \includegraphics[scale = 0.7]{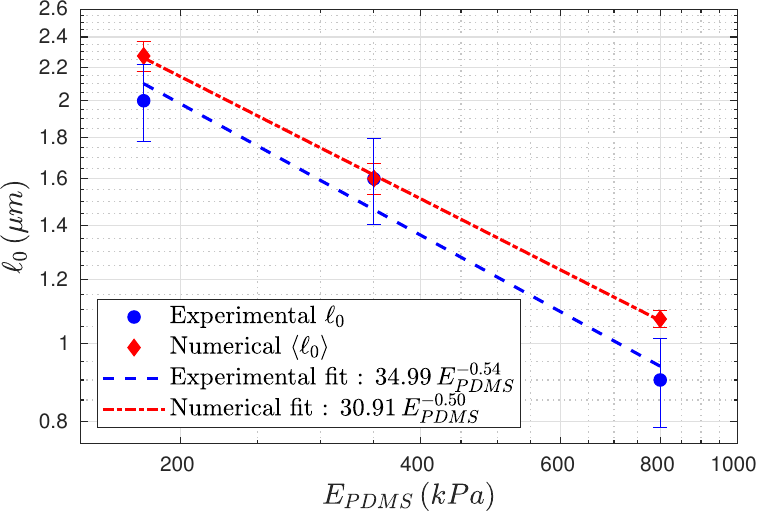}
    \caption{Scaling of the experimental and averaged numerical initial characteristic lengths $\ell_0$ with respect to PDMS Young's modulus $E_{PDMS}$ in the 1D model. The error bars represent the standard deviation.}
    \label{fig:1d_ell0_num_exp}
\end{figure}

Since the initial stochastic perturbation has an influence on the early stage and coarsening phases, the parameters describing the spinodal microstructure are also stochastic. To account for this effect, the characteristics of the spinodal microstructure for each PDMS material are evaluated as the average from ten different computations varying the initial perturbation. Using this procedure we numerically evaluate the initial characteristic length $\langle \ell_0 \rangle$ and time  $\langle \tau \rangle$, where $\langle \bullet \rangle$ denotes the average value of ($\bullet$), for each type of PDMS and we summarize them in Tab.~\ref{tab:1d_ell0_tau}. The obtained numerical values of $\langle \tau \rangle$ indicate that the spinodal decomposition occurs within seconds, in agreement with the qualitative experimental observation in \cite{fernandez-rico_elastic_2023}. 
Also, a comparison of the experimental value in Tab.~\ref{tab:table_PDMS_exp_data} with the numerical ones in  Tab.~\ref{tab:1d_ell0_tau} confirms that, once the model is calibrated to reproduce $\ell_0$ for $E_{PDMS}$ = 350 kPa, the characteristic lengths of the other two materials can be accurately predicted, thus the experimentally observed scaling $\ell_0 \propto 1/\sqrt{E_{EPDMS}}$ is correctly reproduced, as illustrated in Fig.~\ref{fig:1d_ell0_num_exp}. This implies that $m^2/(\gamma E_0)\propto E_{PDMS}$ in \eqref{eq:1d_initial_characteristic_length}.

\begin{table}[]
    \centering
    \begin{tabular}{ccccc}
$E_{PDMS}$ & $\langle \ell_0 \rangle$        & $\langle \tau\rangle$    & $\langle A\rangle$       & $\langle B\rangle$     \\
$(kPa)$    & $(\mu m)$       & $(s)$   & $\left(\mu m \, \sqrt{kPa}\right)$     & $(-)$     \\
$180$      & $2.27$          & $179.50$    & $29.80$    & $0.11$    \\
$350$      & $1.60$          & $52.83$   & $30.47$     & $0.11$   \\
$800$      & $1.07$          & $12.95$     & $29.90$ & $0.11$   
    \end{tabular}
    \caption{\label{tab:1d_ell0_tau} Average values of the numerical initial characteristic lengths and times along with the fitted parameters of the \textit{Ostwald ripening} power law $\ell \, \sqrt{E_{PDMS}} \, = \, A \, (t/\tau)^B$ obtained with the 1D model for different PDMS stiffnesses.}
\end{table}

\begin{figure}[!hbt]
    % \hspace*{-2cm}
    \centering
    \captionsetup{justification=centering}
    \includegraphics[scale = 0.7]{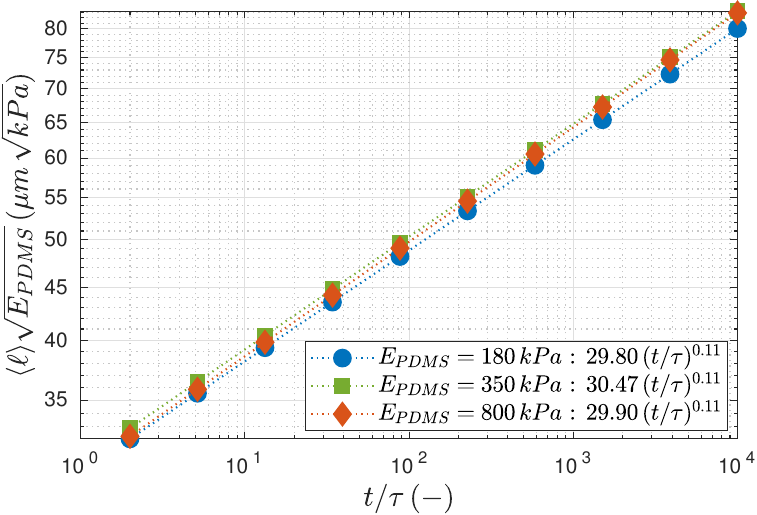}
    \caption{Evolution of the fitted characteristic length during \textit{Ostwald ripening} for the three different PDMS stiffnesses in the 1D model.}
    \label{fig:1d_coarsening_Oswtald}
\end{figure}

In the literature, it is customary to describe the evolution of the characteristic length $\ell$ during \textit{Ostwald ripening} using a power law of the form $\ell(t)=A(t/\tau)^B$. Tab.\ref{tab:1d_ell0_tau} provides the average parameters $\langle A \rangle$ and $\langle B \rangle$ fitted on the computational results for the three PDMS stiffnesses. We observe that the power law coefficient $\langle B \rangle$ remains nearly constant for the three cases, reflecting similar coarsening dynamics. Moreover, Fig.\ref{fig:1d_coarsening_Oswtald} compares the evolution in time of $\langle \ell \rangle\sqrt{E_{PDMS}}$ for the different materials and shows that all curves nicely collapse into one. This demonstrates that the proposed model maintains the scaling $\langle \ell \rangle \propto 1/\sqrt{E_{PDMS}}$ also during the coarsening stage. On the other hand, no coarsening arrest is obtained, motivating the introduction of a long-range interaction energy.

\subsubsection{Case $\alpha > 0$ (Cahn-Oono model coupled with elasticity)} % ----------- Cahn Oono based results  ---------------
\label{subsubsec:1d_cahn_Oono_results}

We now show that adding the long-range interaction \eqref{eq:1d_bulk_energy}$_4$ with $\alpha >0$ allows to arrest coarsening while maintaining the scaling between characteristic length and Young's modulus of the dry polymer. As follows, we perform a parametric study on the coefficient $\alpha$ to better understand its effect on the arrested steady-state characteristic length, denoted as $\ell_D$.

\begin{figure}[!hbt]
    \centering
    \captionsetup{justification=centering}
    \includegraphics[width=\textwidth]{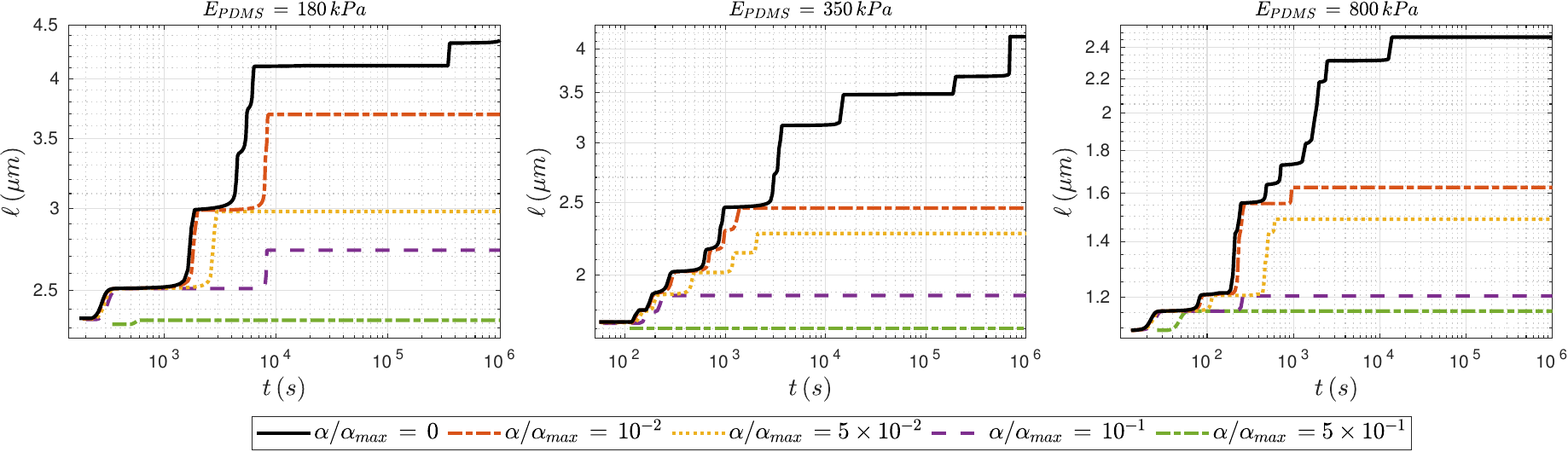}
    \caption{Influence of $\alpha/\alpha_{max}$ on the evolution of the characteristic length for different PDMS stiffnesses in the 1D model.}
    \label{fig:1d_ell_t_alpha_pdms}
\end{figure}

Fig.~\ref{fig:1d_ell_t_alpha_pdms} shows the evolution of the characteristic length for different $\alpha/\alpha_{max}$ ratios for the three {PDMS} types and a single perturbation. Here we always observe the arrest of coarsening with decreasing steady-state characteristic lengths and arrest times for increasing $\alpha$. In particular, for $\alpha/\alpha_{max}=5\times10^{-1}$ the steady-state characteristic length is reached a few seconds after the initial phase separation, while for $\alpha/\alpha_{max}=10^{-2}$ the time increases to almost 2 hours. In the same range of values the steady-state characteristic lengths decrease by up to about 70\%. Fig.~\ref{fig:1d_ell_t_alpha_pdms} suggests that, until arrest is reached, the coarsening dynamics is similar to that obtained with $\alpha=0$ for low values of  $\alpha/\alpha_{max}$ (at least in a time frame up to 10$^6$ s), while for higher values of  $\alpha/\alpha_{max}$ a slower coarsening rate is observed.   

\begin{figure}[!hbt]
    \centering
    \captionsetup{justification=centering}
    \subfloat[]{\includegraphics[scale = 0.7]{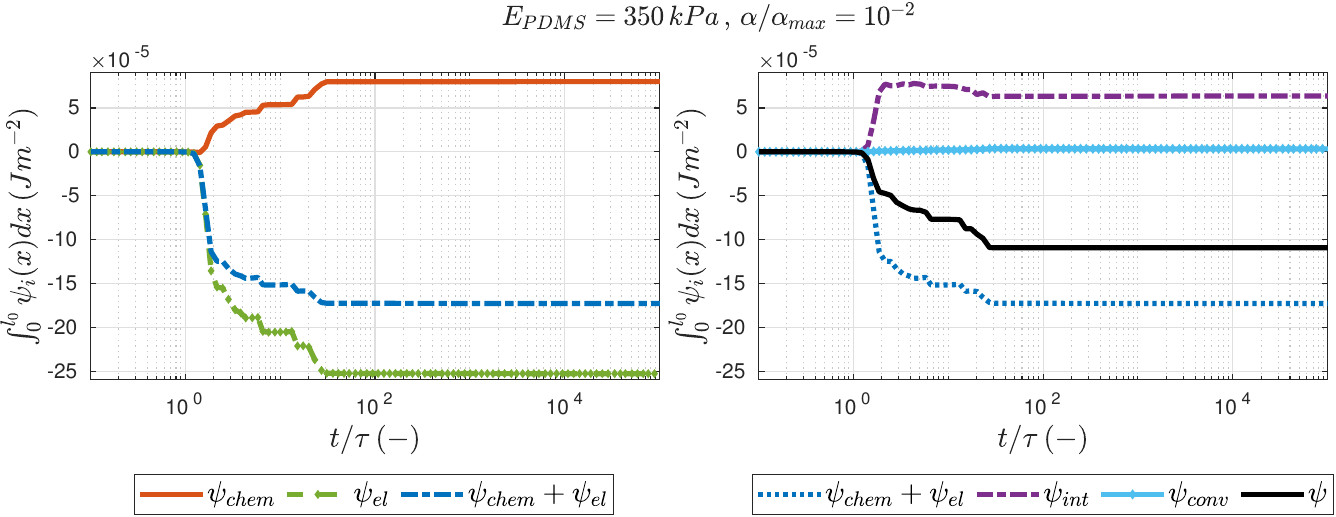}
    \label{fig:1d_energies_350_001}}
    
    \subfloat[]{ \includegraphics[scale = 0.7]{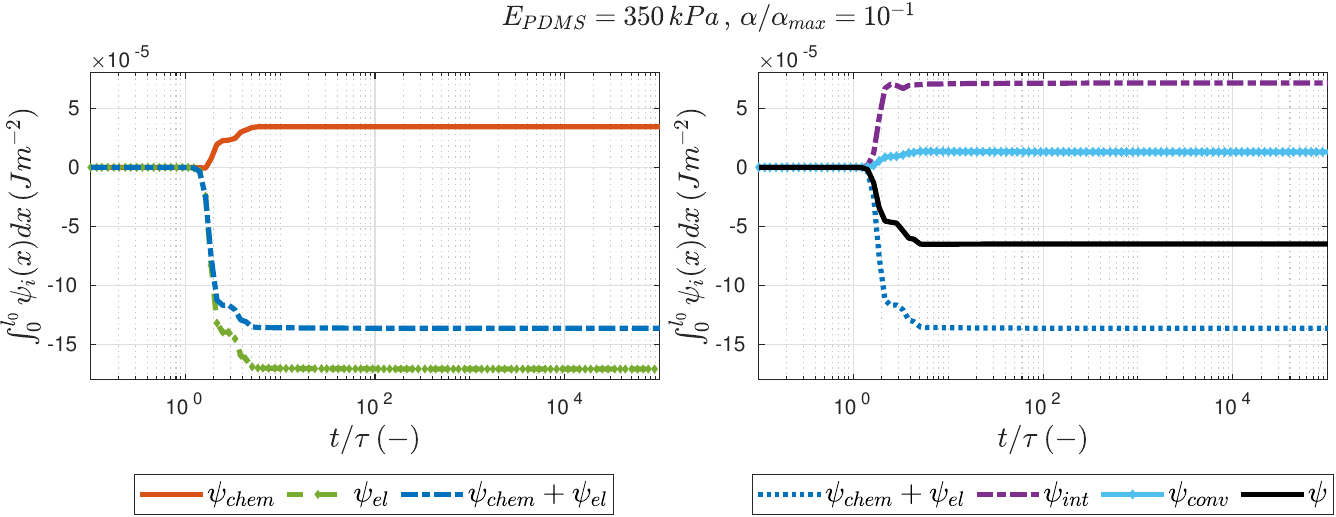}
    \label{fig:1d_energies_350_01} }
    
    \caption{Evolution of the total energy and of its local and nonlocal contributions for the $350 \, kPa$ PDMS in the presence of the long-range interaction term in the 1D model for (a) $\alpha/\alpha_{max}=10^{-2}$ and (b) $\alpha/\alpha_{max}=10^{-1}$.}\label{fig:1d_energies_350}
\end{figure}

Fig. \ref{fig:1d_energies_350} shows the evolution of the various energetic contributions for the PDMS with $E_{PDMS}= 350 \, kPa$ for $\alpha/\alpha_{max}=10^{-2}$ and $10^{-1}$ and clearly illustrates the stabilizing effect of the long-range interactions. The early stages of the spinodal decomposition process are similar to the ones observed in Section \ref{subsubsec:1d_cahn_hilliard_results}. In this case, however, there is an additional positive contribution given by $\int_0^{l_0} \psi_{conv}(x)\,\text{d}x $ that increases at each interface disappearance event\footnote{A positive contribution is present also in the case $\alpha/\alpha_{max}=10^{-2}$ although its magnitude is so small compared to the other contributions that is it not evident in Fig. \ref{fig:1d_energies_350}. In any case, this is already sufficient to arrest the coarsening.}. The competition between decreasing chemical and elastic energy and increasing long-range interaction contribution at some point prevents further coarsening. Hence, the system finds thermodynamic equilibrium (i.e., an energy minimum) with a phase-field variable profile including several segregated clusters of stable phases. Comparing Fig.~\ref{fig:1d_energies_350_001} and \ref{fig:1d_energies_350_01}, we can also appreciate that an increase in $\alpha$ yields stronger long-range interactions, justifying earlier arrest and smaller steady-state characteristic length $\ell_D$. 

\begin{figure}[h!]
    \centering
    \captionsetup{justification=centering}
    \includegraphics[scale = 0.65]{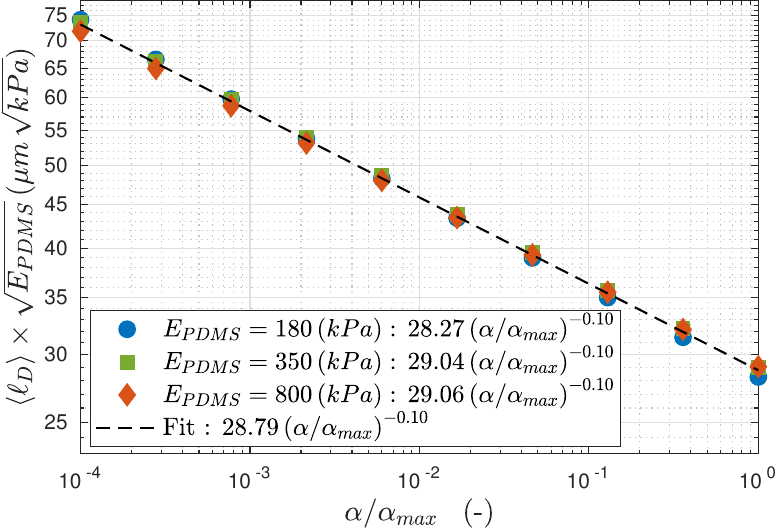}
    \caption{Relation between the steady-state characteristic length $\langle \ell_D \rangle$ scaled by $\sqrt{E_{PDMS}}$  and $\alpha/\alpha_{max}$ obtained from the 1D model.}
    \label{fig:1d_ell_D_alpha_pdms}
\end{figure}

To better characterize the phase separation process, in Fig.~\ref{fig:1d_ell_D_alpha_pdms} we report the relation between average steady-state characteristic length $\langle \ell_D \rangle $ scaled by $\sqrt{E_{PDMS}}$ and $\alpha/\alpha_{max}$, which is well represented by a power law $\ell_D =  C  \, \alpha^{ D }$ similar to the one used for Ostwald ripening (Fig.~\ref{fig:1d_coarsening_Oswtald}). In particular, the value of the exponent in the two cases is very similar but with opposite sign. This property is already observed in \cite{bahiana_cell_1990, liu_dynamics_1989} for the uncoupled {Cahn-Oono} model, for which asymptotic analysis demonstrates that the strict equality $D=-B$ holds. Also, Fig.~\ref{fig:1d_ell_D_alpha_pdms} confirms that the scaling $\langle \ell \rangle \propto 1/\sqrt{E_{PDMS}}$ is preserved when  long-range interactions are accounted for and, in particular, is still valid for the arrested steady-state microstructure. This observation shows that, similarly as in the uncoupled case \cite{politi_dynamics_2014}, the introduction of long-range interactions does not alter considerably the coarsening dynamics.   

\begin{figure}[h!]
    \centering
    \captionsetup{justification=centering}
    \includegraphics[width=\textwidth]{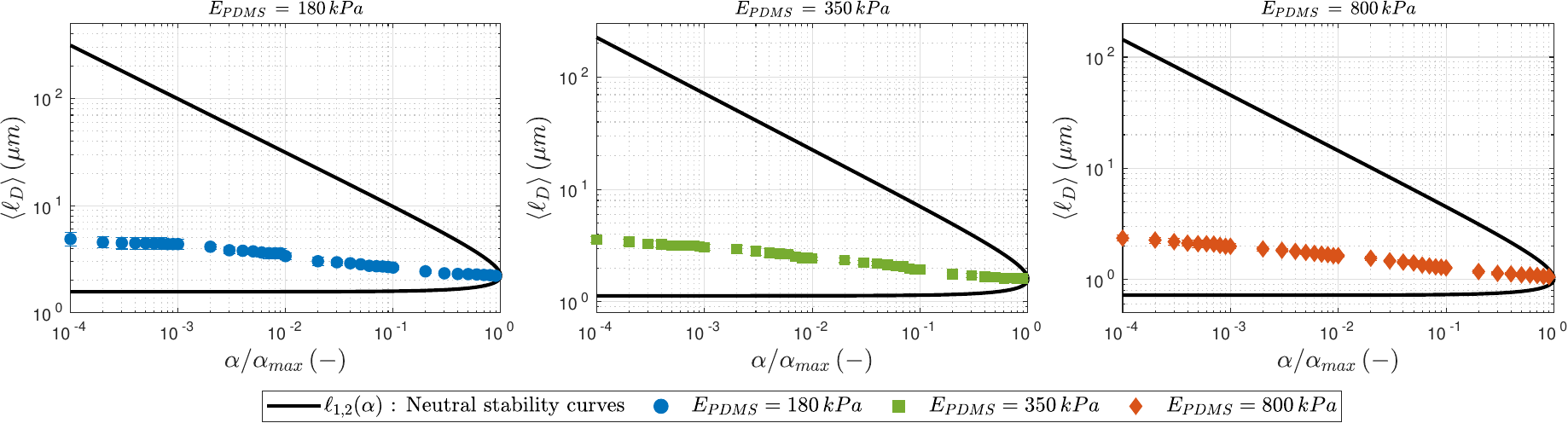}
    \caption{Neutral stability diagram with $\ell_D$ vs. $\alpha/\alpha_{max}$ for the three PDMS stiffnesses in the 1D model.}
    \label{fig:1d_neutral_stability_curve}
\end{figure}

Fig.\ref{fig:1d_neutral_stability_curve} compares the obtained values of $\langle \ell_D \rangle $ for different values of $\alpha/\alpha_{max}$ with the instability region  given by the interval $(\ell_{1},\,\ell_2)$ associated to the wave numbers $(k_1,\,k_2)$ in \eqref{eq:1d_k1_k2_alpha}. As also reported by \cite{politi_dynamics_2014, mcheik_1d_2016} for the uncoupled {Cahn-Oono} framework, $\langle \ell_D \rangle $ remains always inside the instability region. For increasing values of $\alpha/\alpha_{max}$ the instability region reduces and, as $\alpha/\alpha_{max}\rightarrow 1$, it collapses to a point with characteristic length equal to the initial one $\ell_0$. Hence, the coarsening stage does not start. This regime, often referred to as \textit{Eckhaus} scenario of the {Swift-Hohenberg} models \cite{politi_dynamics_2014, benilov_stability_2013}, is characterized by a phase-field variable profile defined by a single wave with a wave number $k=k^*$ given by \eqref{eq:1d_initial_characteristic_length} and illustrated in Fig.~\ref{fig:1d_pulsation_k_cahn_hilliard}. Note that the phase-field pattern obtained in this case for a multi-dimensional setting is different from the standard spinodal pattern, as better illustrated in Section \ref{subsec:2d_numerical_results}. This difference is not appreciable in the 1D setting.

\FloatBarrier

%%%%%%%%%%%%%%%%%%%%%%%% Two-dimensional analysis %%%%%%%%%%%%%%%%%%%%%%
\section{2D model of EMPS}
\label{sec:2d_model} 

In this section, we illustrate 2D governing equations and results from the model in Section \ref{app:final model}, which we then compare with the experimental spatial morphologies emerging from spinodal decomposition. 

\subsection{Problem formulation}    % --- Problem formulation in 1D -----------------
\label{subsec:2d_formulation} 

We now consider the multidimensional model whose free energy density is formulated in Section \ref{app:derivation_coupled_model}, see \eqref{eq:int_energy_final}.

\noindent The elastic part of the energy can be also expressed as 

\begin{equation}
    \label{eq:2d_elastic_energy}
    \psi_{el} =\frac{1}{2} \lambda_0 \, \text{tr}^2(\bm{\tilde \varepsilon})  + G_0 \bm{\tilde \varepsilon} \cdot \bm{\tilde \varepsilon} - m \phi\, \text{tr}(\bm{\tilde \varepsilon}) \,,\quad\text{with}\quad\lambda_0  =\frac{\nu_0 E_0}{(1+\nu_0)(1-\nu_0)}\,\quad \text{and}\quad  G_0  = \frac{E_0}{2(1+\nu_0)}\,,
\end{equation}

\noindent where $\lambda_0$ and $G_0$ are the {Lamé} coefficients of the homogeneous swollen gel. 

The chemical potential and the Cauchy stress tensor read
\begin{equation}
    \hspace{0 mm}
    \label{eq:2d_chemical_potential}
    \tilde \mu =\frac{\partial \psi_{chem}}{\partial \phi} - \nabla_{\!x} \cdot \left( \frac{\partial \psi_{int}}{\partial \nabla_{\!x} \phi} \right) + \frac{\partial \psi_{el}}{\partial \phi} =\gamma \left( \xi\phi + 4\beta\phi^3 \right) - \gamma \kappa \Delta_x \phi -m\,\text{tr}(\bm{\tilde \varepsilon})-\alpha \int_{\mathcal{B}_0} \phi(\bm{s},t)g(\bm{x},\bm{s})d\bm{s}\,,
\end{equation}
\begin{equation}
    \hspace{0 mm}
    \label{eq:stress_strain_2d}
    \bm{\tilde \sigma} = \frac{\partial \psi_{el}}{\partial \bm{\tilde \varepsilon}} =\underbrace{\lambda_0\, \text{tr}(\bm{\tilde \varepsilon})\mathbf{I} + 2G_0\bm{\tilde \varepsilon}\vphantom{\phi \mathbf{I}}}_{\bm{\tilde \sigma}_{el}} \,- \, \underbrace{m\phi \mathbf{I}}_{\bm{\tilde \sigma}_{os}}=\lambda_0\left(\vphantom{\frac{m}{\lambda_0}\phi}\right.\text{tr}(\bm{\tilde \varepsilon})-\underbrace{\frac{m}{\lambda_0}\phi}_{\tilde\varepsilon_{os}} \left. \vphantom{\frac{m}{\lambda_0}\phi} \right)\mathbf{I} + 2G_0\bm{\tilde \varepsilon}  \,,
\end{equation}

\noindent As in Section \ref{subsec:1d_formulation}, in \eqref{eq:stress_strain_2d} we can identify the elastic and the osmotic (volumetric) stress components denoted as $\bm{\tilde \sigma}_{el}$ and $\bm{\tilde \sigma}_{os}$, respectively. Also, from \eqref{eq:stress_strain_2d} we can observe how the osmotic term is proportional to a volumetric eigenstrain $\tilde\varepsilon_{os}\bm{I}$, in turn proportional to the phase-field variable $\phi$. The coefficient $m$ can be related to the expansion coefficient $\Omega$ and the Lamè coefficients through the relationship 
\begin{equation}
    \label{eq:m_coefficient_2D}
        \Omega = \frac{m}{\lambda_0 + G_0}\,,
\end{equation}
\noindent whose derivation is detailed in \ref{app:2d_porous_coefficient}.

The governing equations of mass balance and mechanical equilibrium are 
\begin{equation}
    \label{eq:2d_mass_balance_linear_mom}
    \left\{\begin{split}
        &\frac{\partial \phi}{\partial t}  = -\nabla_{\!x} \cdot\bm{J}, \\
        &\nabla_{\!x} \cdot \bm{\tilde \sigma}  = \bm{0}\,,
    \end{split}\right.
\end{equation}
\noindent where the flux $\bm{J}$ reads
\begin{equation}
    \label{eq:2d_flux}
\bm{J}=-\bm{M} \nabla_{\!x} \tilde\mu= -\bm{M} \left[ \gamma \left( \xi +12\beta\phi^2 \right)\nabla_{\!x} \phi - \gamma \kappa \nabla_{\!x} (\Delta_x\phi) - m\nabla_{\!x} \text{tr}(\bm{\tilde \varepsilon}) - \alpha \int_{\mathcal{B}_0} \phi(\bm{s},t)\nabla_{\!x} g(\bm{x},\bm{s})d\bm{s}  \right] \,.
\end{equation}
\noindent In \eqref{eq:2d_flux}, $\bm{M}$ is the mobility tensor, here assumed to be constant and isotropic, namely $\bm{M} = M \mathbf{I}$. 

Concerning the boundary conditions, we consider a representative area extracted from a 2D version of the S-PDMS specimens in \cite{fernandez-rico_putting_2022} and apply a vanishing average stress condition for the mechanical problem, i.e.
\begin{equation}
    \label{eq:sigma_bc}
\left\langle \bm{\tilde \sigma} \right\rangle =\frac{1}{A_{\mathcal{B}_0}}\int_{\mathcal{B}_0}\bm{\tilde \sigma}\,\text{d}\bm{x} =\bm{0}\,,
\end{equation}
\noindent where $A_{\mathcal{B}_0}$ is the area of the computational domain. For the chemical problem, we adopt periodic boundary conditions for the phase-field variable $\phi$. 

Since the  stress $\bm{\tilde \sigma}$ does not vanish locally in general, it is not possible to determine an analytical expression of the strain $\bm{\tilde \varepsilon}$ from \eqref{eq:2d_mass_balance_linear_mom}, nor to perform a linear stability analysis and to determine an analytical expression for $\alpha_{max}$ as in Section \ref{subsec:1d_linear_stability}.

\subsection{Comparison with Cahn-Larché model}    % ------------ Cahn Larché in 2D --------
\label{subsec:2d_standard_formulation}

Following  \cite{zhu_morphological_2001, garcke_cahnhilliard_2005, voorhees_morphological_1992, orlikowski_large-scale_1999}, the free energy density for the Cahn-Larché model reads

\begin{equation}
    \label{eq:2d_standard_model_free_energy}
    \psi_{CL}( \bm{\tilde \varepsilon}, \phi, \nabla_{\!x} \phi) = \gamma \left( \frac{1}{2}\xi\phi^2 + \beta \phi ^4 \right) + \frac{1}{2}\gamma \kappa |\nabla_{\!x} \phi|^2 + \frac{1}{2} \lambda_0 \, \text{tr}^2 \left( \bm{\tilde\varepsilon} - \Omega \phi \mathbf{I} \right) + G_0 \left( \bm{\tilde \varepsilon} - \Omega \phi \mathbf{I} \right) \cdot \left(\bm{\tilde\varepsilon} - \Omega \phi \mathbf{I} \right)\,.
\end{equation}

\noindent Unlike in the Cahn-Larché models in \cite{zhu_morphological_2001, garcke_cahnhilliard_2005, voorhees_morphological_1992, orlikowski_large-scale_1999}, here we adopt {Lamé} coefficients that are independent of the phase-field variable. The volumetric expansion coefficient $\Omega$ in \eqref{eq:2d_standard_model_free_energy} can be estimated  following \eqref{eq:m_coefficient_2D} in this case as well. Note that, unlike in 1D (see Section \ref{CL-1D}), since the stress does not vanish, the coupling between chemical and mechanical contributions is active.

% ---- Subsection: 2D Numerical Results -----------------
\subsection{Numerical results}  
\label{subsec:2d_numerical_results}  

As in Section \ref{subsec:1d_numerical_results}, we now analyze the results obtained with the proposed model with ($\alpha>0$) or without ($\alpha=0$) long-range interactions and we compare the results to those of the Cahn-Larché model in Section \ref{subsec:2d_standard_formulation}. We solve the coupled problem \eqref{eq:2d_mass_balance_linear_mom} with the finite element method. We consider a square domain within a purely 2D setting (i.e. ignoring the third dimension) with dimensions 40$\times$40 $\mu m^2$ and discretize it with 400$\times$400 bilinear quadrilateral elements. Further computational aspects are summarized in Section \ref{app:solution_strategy}.

\subsubsection{Parameter calibration}  % ---------parameters Calibration

Since a linear stability analysis is not feasible in the multi-dimensional case, we cannot analytically calibrate the parameters $\gamma, \, \beta, \, \kappa$ as in Section \ref{subsubsec:1d_calibration}.  Instead, following again the ideas in \ref{app:1d_calibration}, we perform a set of numerical computations to obtain the curves relating $\ell_0$ with $\gamma$ and $\kappa$, and $\beta$ with the binodal points $(\phi^1_b,\,\phi^2_b)$. Also in this case, the parameters $\gamma$ and $\kappa$ are calibrated for the material with $E_{PDMS}=350$ kPa and then extended to the other materials. The influence of $\gamma, \, \beta, \, \kappa$ on $\ell_0$ and $(\phi^1_b,\,\phi^2_b)$ remains qualitatively the same as in the 1D model. In addition to the mechanical properties in Tab.~\ref{tab:table_PDMS_exp_data}, we assume here a constant Poisson's ratio $\nu_0 = 0.45$ for the S-PDMS system.
%based on the values reported in \cite{dogru_poissons_2018}. 
The parameters adopted for the 2D computations are summarized in Tab.~\ref{tab:2d_table_num_data}.

\begin{table}[h!]
    \centering
    \begin{tabular}{ccccccc}
$E_{PDMS}$ & $\nu_0$ & $\gamma$    & $\xi$ & $\beta$          & $\kappa$          &  $m$\\
$(kPa)$    & (-) & $(Jm^{-3})$ & $(-)$           & $(-)$            & $(m^2)$       & $(kPa) $   \\
$180$ & 0.45     & $10^4$      & $-0.12$         & $2.05 \, 10^{1}$ & $4.60\, 10^{-14}$ & $2.85 \, 10^{1}$   \\
$350$ & 0.45     & $10^4$      & $-0.12$         & $3.94 \, 10^{1}$ & $4.60\, 10^{-14}$ & $5.81 \, 10^{1}$  \\
$800$ & 0.45     & $10^4$      & $-0.12$         & $1.02 \, 10^{2}$ & $4.60\, 10^{-14}$ & $1.41 \, 10^{2}$   
\end{tabular}
\caption{\label{tab:2d_table_num_data} Parameters adopted for the 2D  computations using the proposed model.}
\end{table}

For the Cahn-Larché model, we retain the values of the parameters in Tab.~\ref{tab:table_PDMS_exp_data} and \ref{tab:2d_table_num_data} except for the interface parameters $\kappa$ and $\beta$. Following the same approach used for the proposed model,  $\kappa$ is adjusted so that the experimental and numerical initial characteristic lengths of the PDMS with $E_{PDMS}= 350 \, kPa$ coincide, while $\beta$ is tuned for all materials to obtain a composition at the binodal points $(\phi_b^1,\,\phi^2_b)=(-0.1, \, 0.1)$ as also assumed in \ref{app:1d_calibration}. The obtained parameters are summarized in Tab.~\ref{tab:2d_table_num_data_CL}.

\begin{table}[h!]
\centering
\begin{tabular}{cccc}
$E_{PDMS}$ & $\beta$          & $\kappa$    & $\Omega$   \\
$(kPa)$    & (-) & $(m^2)$  &    $(-)$       \\
$180$ & $3.70$     & $1.82 \, 10^{-15}$      & $0.23$    \\
$350$ & $4.14$     & $1.82 \, 10^{-15}$      & $0.24$  \\
$800$ & $4.38$     & $1.82 \, 10^{-15}$      & $0.26$    
\end{tabular}
\caption{\label{tab:2d_table_num_data_CL} Parameters adopted for the 2D computations using the Cahn-Larché model.}
\end{table}

\subsubsection{Case $\alpha = 0$ (Cahn-Hilliard model coupled with elasticity)}  %----- Cahn Hilliard elasticity based results 2D ------------
\label{subsubsec:2d_cahn_hilliard_results}

\begin{figure}[!hbt]
    \hspace*{-1.0cm}
    \vspace*{0.1 cm}
    \centering
    \captionsetup{justification=centering}
    \includegraphics[trim={1cm 1.65cm 0cm 0cm}]{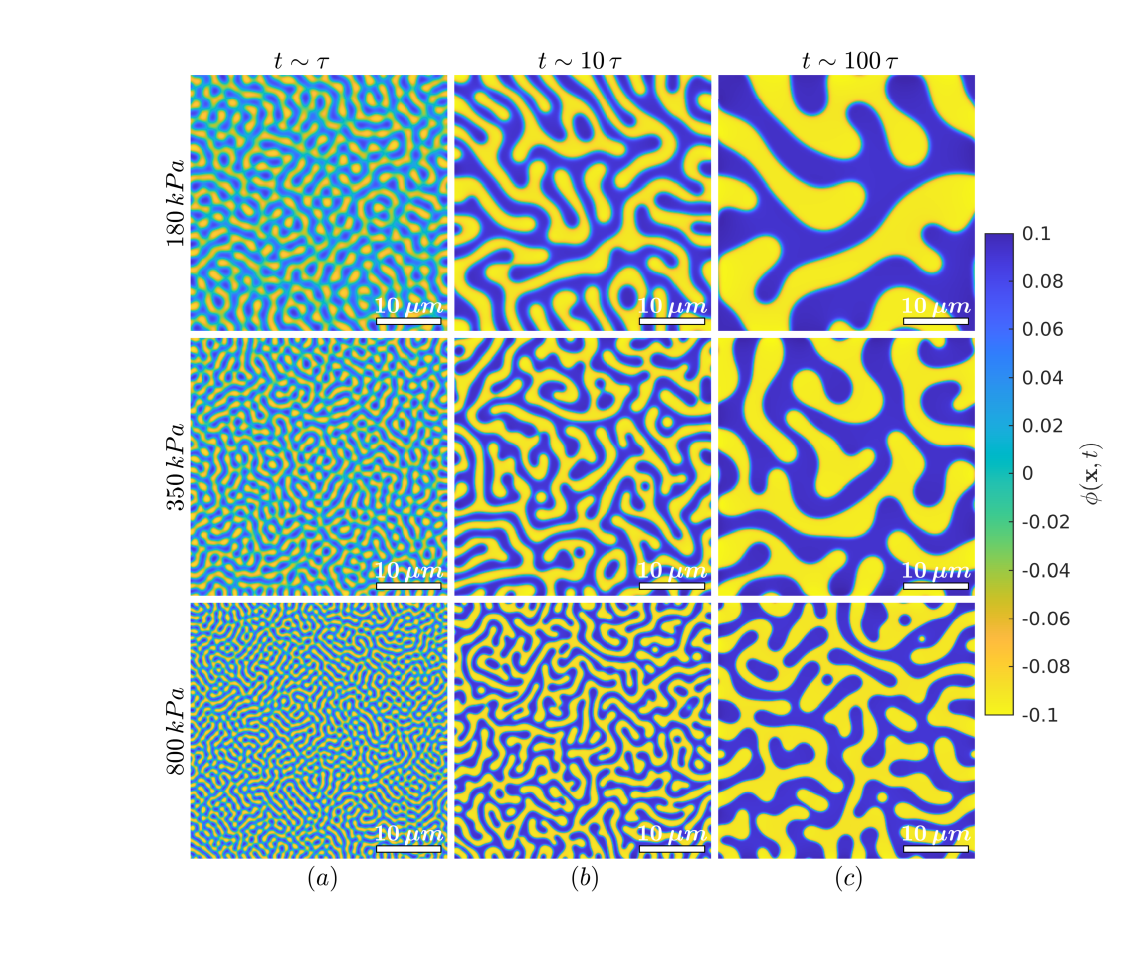}
    \caption{$\phi(\bm{x},t)$ profile at $t\sim \tau \, (a)$, $t \sim 10\tau \, (b)$ and $t \sim 100\tau \, (b)$ for different PDMS stiffnesses (rows) in the 2D model.}
    \label{fig:2d_PDMS_morphologies}
\end{figure}

\begin{figure}[!hbt]
    \hspace*{-0.5cm}
    \centering
    \captionsetup{justification=centering}
    \includegraphics[trim={0cm 2.25cm 0cm 0cm}]{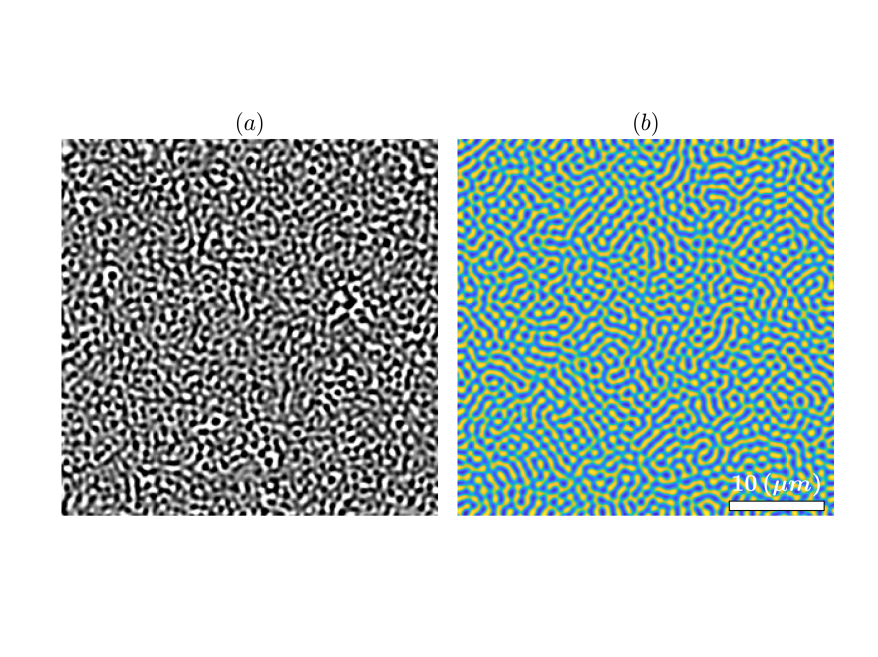}
    \caption{Comparison of the experimental $(a)$ and the numerical $(b)$ initial morphologies of the mixtures with $E_{PDMS}=800 \, (kPa)$.}
    \label{fig:2d_pattern_comparison}
\end{figure}

Fig.~\ref{fig:2d_PDMS_morphologies} shows the spinodal decomposition morphologies at different stages of the coarsening process. At the {early} stage, i.e. for $t \sim \tau$ (Fig.~\ref{fig:2d_PDMS_morphologies}a), the homogeneous mixture phase separates into a bicontinuous channel-like structure. Its morphology is very similar to the experimental one, as illustrated in Fig.~\ref{fig:2d_pattern_comparison} for $E_{PDMS}= 800 \, kPa$. 

\begin{figure}[h!]
    \centering
    \captionsetup{justification=centering}
    \includegraphics[scale = 0.8]{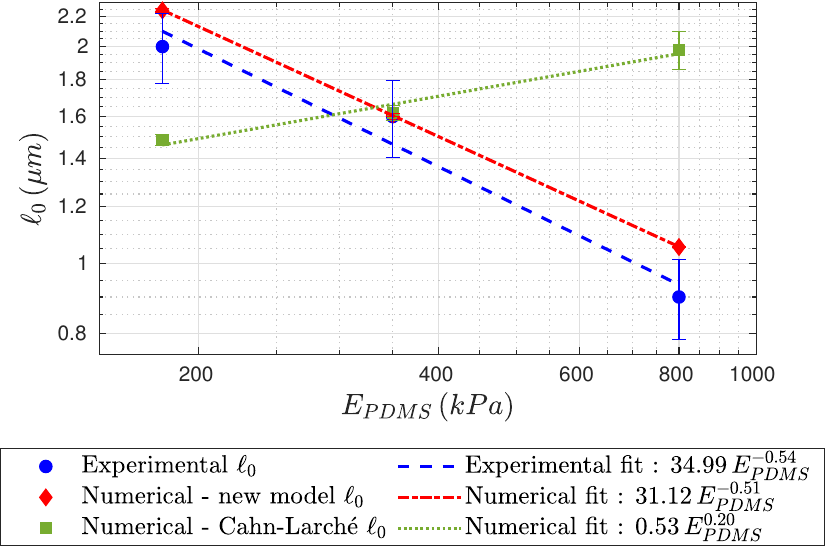}
    \caption{Scaling of $\ell_0$ with respect to $E_{PDMS}$ in the 2D model.}
    \label{fig:2d_ell_0}
\end{figure}

Since in 2D the initial perturbation has less influence on the final results than in 1D, here the characteristics of the spinodal structures are evaluated based on a single computation. Tab. \ref{tab:2d_lambda_exp_num} summarizes the computed  initial characteristic lengths. While the parameters $\kappa$ and $\beta$ are calibrated for  $E_{PDMS}=350\ kPa$, the results for $E_{PDMS}=180\ kPa$ and $800\ kPa$ are in very good agreement with the experimental values (Tab.~\ref{tab:table_PDMS_exp_data}). As a result, the experimentally observed scaling $\ell_0 \propto 1/\sqrt{E_{PDMS}}$ is correctly predicted by the model also in 2D, as illustrated in Fig.~\ref{fig:2d_ell_0}. The numerical initial characteristic times are all below 35 seconds, suggesting a rapid spinodal decomposition as qualitatively observed also in the experiments. 

In Fig.~\ref{fig:2d_ell_0} we also report the initial characteristic lengths obtained from the Cahn-Larché model. The corresponding trend shows a slight increase with increasing PDMS stiffness, which goes against the experimental evidence. As a result, the obtained length is similar to the experimental observation only for the material used for calibration (i.e., for $E_{PDMS}=350\ kPa$), and significantly different for the other materials. 

\begin{table}[]
\centering
\begin{tabular}{cccc}
$E_{PDMS}$ & $\ell_0$        & $\tau$      \\
$(kPa)$    & $(\mu m)$       & $(s)$       \\
$180$      & $2.24$          & $34.84$         \\
$350$      & $1.61$          & $10.26$         \\
$800$      & $1.06$          & $2.63$        
\end{tabular}
\caption{\label{tab:2d_lambda_exp_num} Numerical initial characteristic lengths and times for different PDMS stiffnesses in the 2D model.}
\end{table}

\begin{figure}[!htb]
    \centering
    \captionsetup{justification=centering}
    \includegraphics[scale = 0.8]{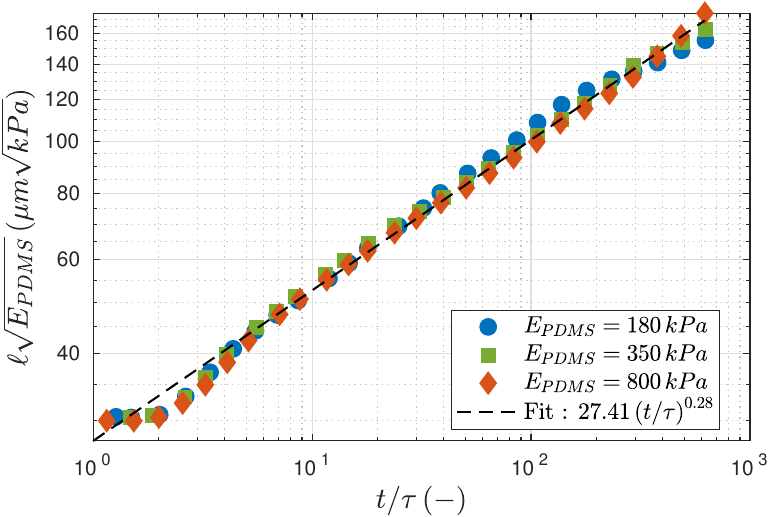}
    \caption{Evolution of the scaled characteristic length for different PDMS stiffnesses for a given initial perturbation in the 2D model.}
    \label{fig:2d_ell_t_evolution}
\end{figure}

\begin{figure}[!htb]
    \centering
    \captionsetup{justification=centering}
    \includegraphics[scale = 0.75]{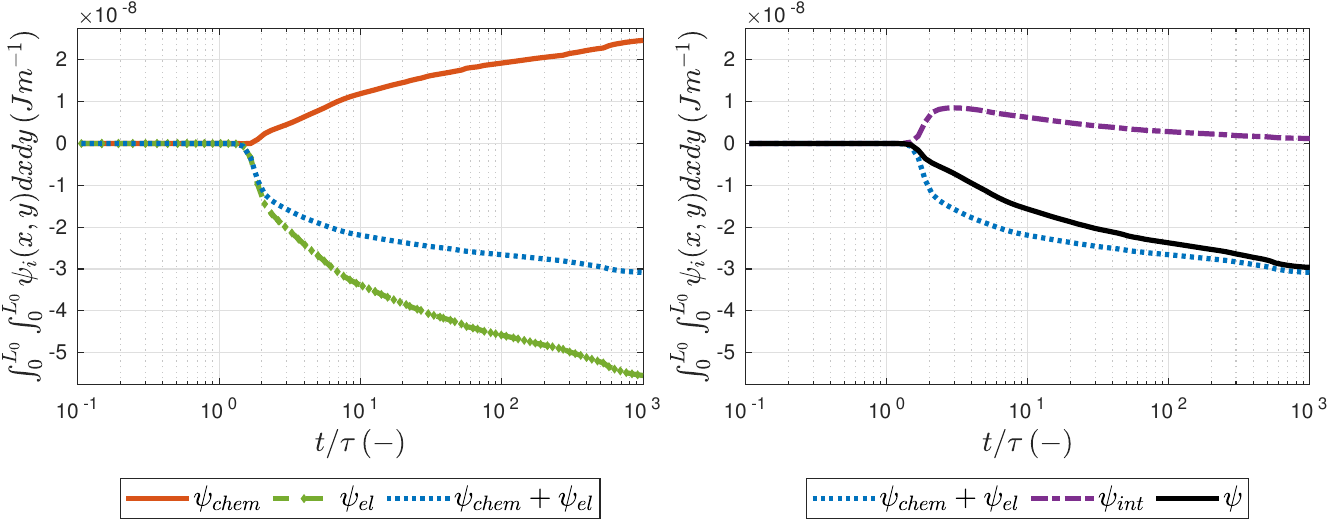}
    \caption{Evolution of the total energy and of its local and nonlocal contributions for the mixture with $E_{PDMS}=350 kPa$ for a given initial perturbation in the 2D model.}
    \label{fig:2d_PDMS_energies}
\end{figure}

After the early stage, the coarsening phase starts where the characteristic length of the channel-like spinodal structures steadily increase for all materials  as illustrated in Figs.~\ref{fig:2d_PDMS_morphologies}b-c for $t\sim 10\tau$ and $100\tau$, respectively. The time evolution of the characteristic lengths scaled by $\sqrt{E_{PDMS}}$ is reported in Fig.~\ref{fig:2d_ell_t_evolution}, where the initial  plateau  for $t/\tau \leq 1$ represents the early-stage phase decomposition, while the following branch represents the coarsening stage. We note that, as in 1D, the coarsening of the spinodal structures is not arrested. This is consistent with the results in Fig.~\ref{fig:2d_PDMS_energies}, where we plot the evolution of the different energetic terms during the initial and coarsening stages of spinodal decomposition. As in the 1D case, the interfaces give a positive contribution toward  the total energy, thus the system evolves by reducing the number of interfaces. The dissolution of interfaces results in larger domains with homogeneous composition, which are also associated to lower chemical and elastic energies. 

Fig.~\ref{fig:2d_ell_t_evolution} also demonstrates  that the curves for the different values of $E_{PDMS}$ are almost superimposed, confirming that the scaling of the initial length $\ell_0 \propto 1/\sqrt{E_{PDMS}}$ continues to be valid during the coarsening stage. Also here, the evolution of the characteristic length during coarsening follows the {Ostwald} ripening law $\ell(t) = At^B$, where the fitted coefficients are summarized in Tab.\ref{tab:2d_Ostwald_ripening_law}. The obtained coarsening rate is $B \sim 0.28$, a value lower than $B \sim 0.33$ obtained adopting the standard uncoupled {Cahn-Hilliard} model in a multi-dimensional setting \cite{chakrabarti_late-stage_1993, konig_two-dimensional_2021, dai_computational_2016}. The slower coarsening rate is due to the osmotic term $\bm{J}_{os}=m\,M \,\nabla_{\!x} \text{tr}(\bm{\tilde \varepsilon})$ in \eqref{eq:2d_flux}, which promotes diffusion in the direction of increasing total volumetric strains. In case of free deformations with vanishing average stress, the spatial distribution of the volumetric strains in \eqref{eq:stress_strain_2d} is mainly determined by the concentration $\phi$ through the osmotic term $\varepsilon_{os}\bm{I}$, which increases for an increasing concentration. Thus, the direction of the osmotic flux $\bm{J}_{os}$ is opposite to that of the Cahn-Hilliard diffusive term $\bm{J}_{CH}=-M \gamma \left( \xi +12\beta\phi^2 \right)\nabla_{\!x} \phi$, leading to a global reduction of the coarsening rate. Similar effects are also observed in the Cahn-Larché model when the Lamè coefficients depend on the phase-field variable \cite{nishimori_pattern_1990, zhu_morphological_2001}. In this case, the observed coarsening rates depend on the contrast between the elastic constants at the binodal points and usually oscillate within $B \sim 0.2-0.3$.

\begin{table}[]
    \centering
    \begin{tabular}{ccc}
$E_{PDMS}$ & $A$                                            & $B$                           \\
$(kPa)$    & $\left(\mu m \, \sqrt{kPa}\right)$             & $(-)$                         \\
$180$      & $27.62$                                        & $0.28$                        \\
$350$      & $27.90$                                        & $0.28$                        \\
$800$      & $26.71$                                        & $0.28$             
    \end{tabular}
    \caption{\label{tab:2d_Ostwald_ripening_law} \textit{Ostwald ripening} power law $\ell \, \sqrt{E_{PDMS}} \, = \, A \, (t/\tau)^B$ during the coarsening stage in the 2D model.}
\end{table}

\FloatBarrier
\subsubsection{Case $\alpha > 0$ (Cahn-Oono model coupled with elasticity)}   %--------- Cahn Oono elasticity 2D based results ----
\begin{figure}[!hbt]
    \hspace*{-1.0cm}
    \centering
    \captionsetup{justification=centering}
    \includegraphics[scale = 0.6]{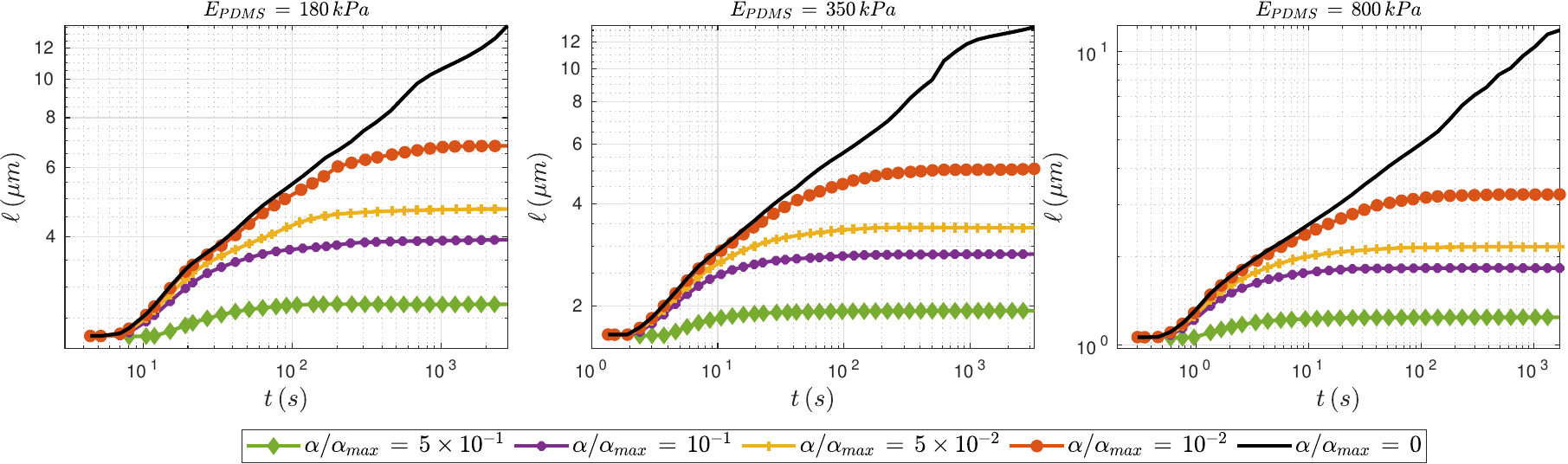}
    \caption{Influence of $\alpha/\alpha_{max}$ on the evolution of the characteristic length $\ell$ for different PDMS stiffnesses in the 2D model.}
    \label{fig:2d_ell_t_alpha_pdms}
\end{figure}

\label{subsubsec:2d_cahn_Oono_results} Tab.\ref{tab:2d_alpha_max_num} summarizes the values of $\alpha_{max}$, which are obtained numerically by performing  computations for increasing values of $\alpha$ until phase separation is suppressed. Fig.~\ref{fig:2d_ell_t_alpha_pdms} plots the evolution of the characteristic length for different $\alpha/\alpha_{max}$ ratios. The obtained results essentially confirm what already observed for the 1D model (Section \ref{subsubsec:1d_cahn_Oono_results}). Initially, the phase separation process shows a trend similar to that obtained for $\alpha=0$, with an early stage characterized  by a sub-horizontal branch followed by a coarsening phase. In this case, however, the coarsening rate first increases and then starts decreasing, so that ultimately a steady-state spinodal characteristic length $\ell_D$ is attained. 

\begin{table}[!hbt]
    \centering
    \begin{tabular}{cc}
$E_{PDMS}$ & $\alpha_{max}$ \\
$(kPa)$    & $(J\mu m^{-5})$    \\
$180$      & $2.6 \, 10^{-14}$ \\
$350$      & $1.0 \, 10^{-13}$ \\
$800$      & $6.1 \, 10^{-13}$         
    \end{tabular}
    \caption{\label{tab:2d_alpha_max_num} Long-range coefficient limit $\alpha_{max}$ for the different PDMS stiffnesses in the 2D model.}
\end{table}

\begin{figure}[!hbt]
    \hspace*{-1cm}
    \centering
    \captionsetup{justification=centering}
    \includegraphics[scale = 0.75]{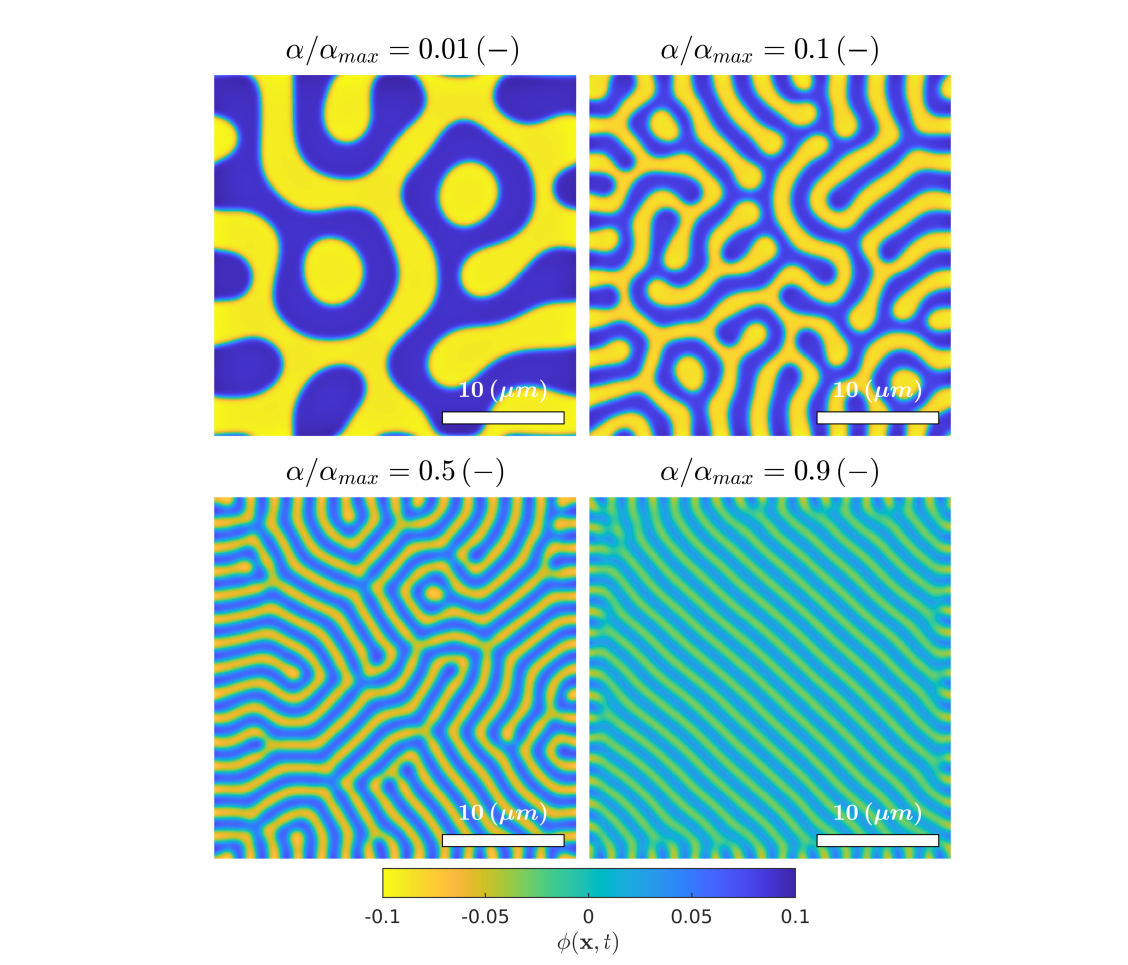}
    \caption{Arrested microstructures for different  ratios $\alpha/\alpha_{max}\geq0.01$ for $E_{PDMS}=350 \, kPa$ in the 2D model.}
    \label{fig:2d_pattern_350}
\end{figure}

\begin{figure}[!hbt]
    \hspace*{-1cm}
    \centering
    \captionsetup{justification=centering}
    \includegraphics[scale = 0.75]{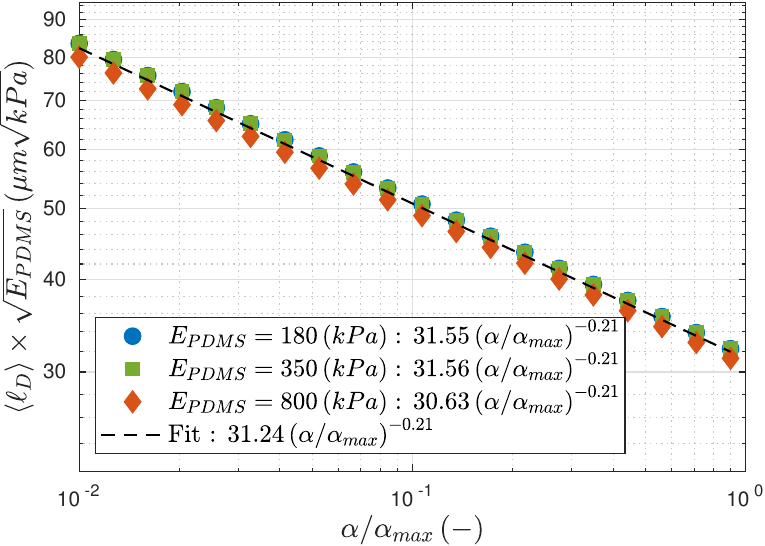}
    \caption{Evolution of the scaled steady-state characteristic length $\ell_D\times\sqrt{E_{PDMS}}$ for different $\alpha/\alpha_{max}$ ratios and for different PDMS stiffnesses in the 2D model.}
    \label{fig:2d_ell_D_scaling_alpha}
\end{figure}

For values of $\alpha/\alpha_{max}\le0.1$ the early stage shows initial characteristic times $\tau$ and coarsening rates before the arrest similar to the case with $\alpha=0$, while for higher values of $\alpha/\alpha_{max}$ we observe a delayed phase separation initiation and a slower coarsening dynamic. An increase of the value of $\alpha/\alpha_{max}$ also leads to a decrease of the steady-state characteristic length $\ell_D$ and of the time needed to reach it (Figs. \ref{fig:2d_ell_t_alpha_pdms} and \ref{fig:2d_pattern_350}). This can be better appreciated in Fig.~\ref{fig:2d_ell_D_scaling_alpha} which illustrates the evolution of the scaled arrested characteristic length $\ell_D \sqrt{E_{PDMS}}$ as a function of $\alpha/\alpha_{max}$. The results demonstrate that the scaling $\ell_D \propto \sqrt{E_{PDMS}}$ is preserved  also in the 2D setting and for any $\alpha/\alpha_{max}$ ratio, confirming the conclusions drawn for the 1D model. 

The 2D model allows to better appreciate the distinction between the {Eckhaus} scenario and the standard chaotic spinodal microstructure regime mentioned in Section \ref{subsubsec:1d_cahn_Oono_results}. For $\alpha/\alpha_{max}\simeq 1$, the range of wave numbers $[k_1,\,k_2]$ associated with spontaneous phase separation degenerates to a point, i.e. $k_1=k_2$. Hence,  the phase-field variable profile becomes a modulated pattern described by a single sinusoidal wave and the arrested characteristic length $\ell_D$ is close to $\ell_0$ leading to the so-called {Eckhaus} regime  (Fig.~ \ref{fig:1d_neutral_stability_curve}). As a consequence, the emerging pattern displays stripes (e.g., Fig.\ref{fig:2d_pattern_350} for $\alpha/\alpha_{max}=0.9$) unlike the classical spinodal microstructure obtained for lower $\alpha/\alpha_{max}$ ratios. For intermediate values of $\alpha/\alpha_{max}$ (e.g., $\alpha/\alpha_{max}=0.5$), the steady-state mixtures exhibit a mix between a spinodal and a striped pattern.

\subsubsection{Reproducing the experimental results} % ----- Reproducing the experimental results section ---
In the previous examples, we calibrated the models so that the  observed characteristic spinodal length in \cite{fernandez-rico_elastic_2023} is equal to the initial one. This assumption is needed to allow for a direct comparison between the analyzed models since they all reproduce the early stage of the spinodal decomposition and hence have an initial characteristic length. The proposed model, however, is also able to reproduce a coarsening arrest and, hence, a steady-state characteristic length. Since we do not know whether the length measured in \cite{fernandez-rico_elastic_2023} is the initial or a partially coarsened one (Section \ref{sec:Introduction}), here we show that the proposed model can be calibrated to reproduce the experimentally measured characteristic length as a steady-state length.

\begin{figure}[!hbt]
    \hspace*{-1cm}
    \centering
    \captionsetup{justification=centering}
    \includegraphics[scale = 0.7]{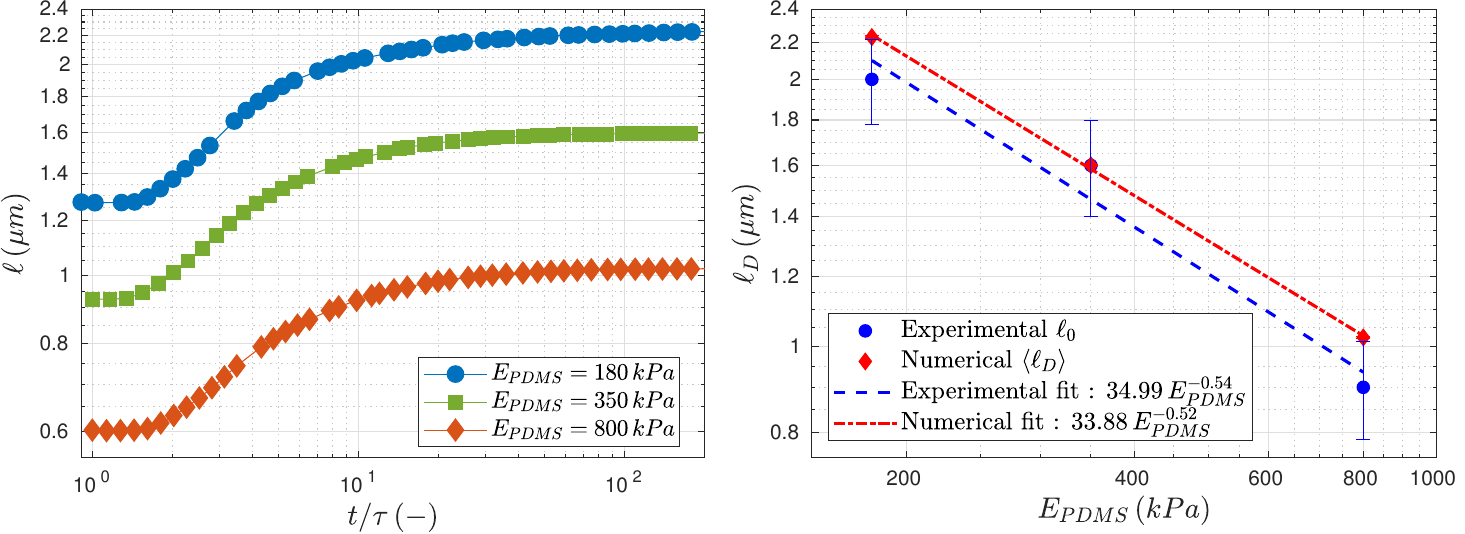}
    \caption{Evolution of the characteristic lengths for different PDMS stiffnesses and comparison of the steady-state characteristic length against the observed experimental one for $\alpha/\alpha_{max}=0.1$ and $\kappa = 1.46\,10^{-14} \, (m^2)$.}
    \label{fig:2d_ell_t_alpha_ell_D_ell_0}
\end{figure}

\begin{figure}[!hbt]
    \hspace*{-0.5cm}
    \centering
    \captionsetup{justification=centering}
    \includegraphics[trim={0cm 2.25cm 0cm 0cm}]{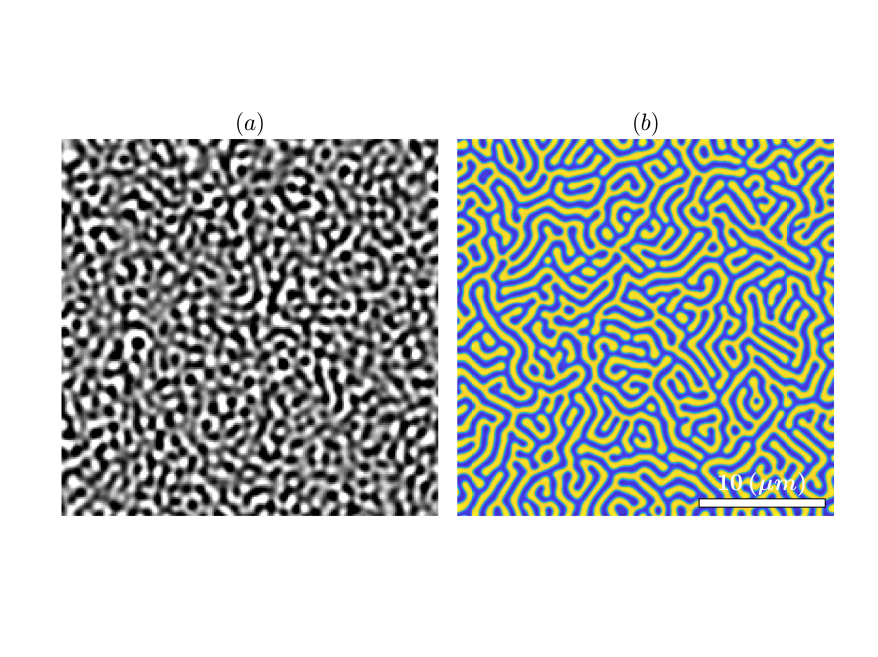}
    \caption{Comparison of the experimental $(a)$ and the numerical $(b)$ arrested morphologies of the $800 \, (kPa)$ PDMS mixture for $\alpha/\alpha_{max}=0.1$ and $\kappa = 1.46\,10^{-14} \, (m^2)$.}
    \label{fig:2d_pattern_comp_alpha_0_1}
\end{figure}

We start by considering that the experimental tests in \cite{fernandez-rico_putting_2022} report a random spinodal pattern rather than a stripe-like arrangement. Hence, in this comparison we focus on values of $\alpha/\alpha_{max}$ sufficiently low to exclude the Eckhaus scenario, namely $\alpha/\alpha_{max} = 10^{-1}$. Also, considering that the steady-state characteristic length $\ell_D$ is always larger than (or, at most, equal to) the initial one $\ell_0$, we calibrate the model by reducing $\ell_0$ so that the coarsening is arrested exactly for $\ell_D$ equal to the experimentally measured length. To this end, we start by  noting that $\ell_0$ is related to $\gamma$ and $\kappa$. We keep $\gamma = 10^4 \,(Jm^{-3})$ to retain the same relative importance of the chemical part compared to the mechanical one that we had in the previous calibration. Therefore, we modify $\kappa$, which now takes the value $\kappa = 1.46\,10^{-14} \, (m^2)$. Once again, we perform the calibration for one PDMS material. We report in Fig.\ref{fig:2d_ell_t_alpha_ell_D_ell_0} the evolution in time of the characteristic lengths. As expected, the numerical steady-state characteristic length $\ell_D$ coincides with the experimentally measured one. Also, the arrest occurs within a time frame of the order of a few minutes, compatible with the experimental observations. The numerical arrested morphology keeps a channel-like structure close to the experimental one, see Fig.\ref{fig:2d_pattern_comp_alpha_0_1}.

\FloatBarrier

%%%%%%%%%%%%%%%%%%%%%%%% Conclusions %%%%%%%%%%%%%%%%%%%%%%
\section{Conclusions}
\label{sec:conclusions}
We proposed a novel phase-field model of EMPS in polymer gels. The model is based on a free energy density depending on a strain tensor, which represents the incremental infinitesimal strain with respect to the swollen homogeneous state of the gel, on a phase-field parameter, related to the volume fraction of the solvent in the polymer gel, and on its gradient. The coupling between the strain and the phase-field parameter is naturally obtained from a derivation that starts from an  entropic elastic energy density combined with the assumption of weak compressibility. Second-order approximation of this energy around the swollen state, similarly as in the linearization of the theory in \cite{hong_theory_2008} performed in \cite{bouklas_swelling_2012, bouklas_nonlinear_2015}, leads to a poroelastic formulation where the coupling energetic term can be interpreted as the osmotic work of the solvent within the polymer matrix. Unlike in \cite{hong_theory_2008,bouklas_swelling_2012, bouklas_nonlinear_2015}, the free energy density of our model also incorporates a chemical and an interface contribution of Cahn-Hilliard type, which enable the model to predict phase separation and spinodal decomposition. The resulting model gives significantly different predictions than the Cahn-Larché model, classically used to describe spinodal decomposition coupled with elasticity effects \cite{zhu_morphological_2001, garcke_cahnhilliard_2005, voorhees_morphological_1992, orlikowski_large-scale_1999}. In particular, our model is able to reproduce the experimentally observed scaling of the characteristic length of the spinodal structure with the stiffness of the dry polymer. This scaling is observed in the initial characteristic length (the one at the onset of spinodal decomposition) and continues to hold during the coarsening stage. Additionally, when extended with a Cahn-Oono term accounting for nonlocal interactions, our model is able to predict arrest of the spinodal decomposition, with a final characteristic length again depending on the stiffness of the dry polymer according to the experimentally observed scaling. 

\clearpage
%%%%%%%%%%%%%%%%%%%%%%%% Appendix %%%%%%%%%%%%%%%%%%%%%%%%%%
\appendix
\section{Expansion of the elastic energy density in \eqref{eq:elastic_flory_expanded_MB} around the swollen state}
\setcounter{figure}{0}
\setcounter{table}{0}

\label{app:linear_elast}
Let us recall for convenience the elastic energy density \eqref{eq:elastic_flory_expanded_MB}

\begin{equation}
    \label{eq:elastic_flory_expanded}
    \Psi_{el}(\bm{F}, \Phi) = \underbrace{\frac{1}{2}Nk_BT\left[ \mathrm{tr}(\bm{C}) - 3 -2 \log(J) \right]}_{\Psi_{ent}} + \underbrace{\frac{K}{2}\left[J- (1+\Omega \Phi) \right]^2}_{\Psi_{bulk}}.
\end{equation}

%\noindent From 
%\ref{eq:elastic_flory_expanded}, the \textit{Cauchy} stress tensor $\bm{\sigma}$ is obtained as
%\begin{equation}
    %\label{eq:cauchy_stress_flory}
    %\bm{\sigma}  = \frac{1}{J} \frac{\partial \Psi}{\partial \bm{F}} \bm{F}^T = \frac{1}{J}Nk_BT\left(\bm{F}\bm{F}^T - \bm{I} \right) + K\left[J- (1+\Omega \Phi) \right]\bm{I}\,.
%\end{equation}
%\noindent 
%\noindent In the intermediate configuration we have
%\begin{equation}
    %\label{eq:cauchy_stress_flory_int}
    %\bm{\sigma}_0   = \frac{1}{J_0}Nk_BT\left(\bm{F}_0\bm{F}_0^T - \bm{I} \right) + K\left[J_0- (1+\Omega \Phi_0) \right]\bm{I}=\bm{0}\quad \Rightarrow \quad J_0-(1+\Omega\Phi_0)=-\frac{Nk_B T}{K J_0}(\delta_0^2-1)\,,
%\end{equation}
%\noindent where we used \eqref{eq:def_grad_0} and the stress-free assumption at $t=t_0$.
We now expand \eqref{eq:elastic_flory_expanded} around the homogeneous swollen state up to second order. To this end we introduce the following expansions around the intermediate configuration $\mathcal{B}_0$:
\begin{equation}
\begin{split}
    \label{eq:linearized_kinematics}
        \nabla_{\!X} \tildbm{u} = \nabla_{\!x} \tildbm{u} \bm{F}_0\,
        , \quad \bm{F} &= \bm{F}_0 + \nabla_{\!x}\tildbm{u}\bm{F}_0
\end{split}
\end{equation}
from which
\begin{equation}
    \label{eq:C_exp}
\tr\left(\bm{C}\right)=\tr\left(\bm{F}^T\bm{F}\right)=\tr\left[\left(\bm{F}_0+\nabla_{\!x} \tildbm{u}\bm{F}_0\right)^T\left(\bm{F}_0+\nabla_{\!x} \tildbm{u}\bm{F}_0\right)\right]=3\delta_0^2+2\delta_0^2\tr\left(\tildbm{\varepsilon}\right)+\delta_0^2\tr\left(\tildbm{\varepsilon}^2\right)-\delta_0^2\tr\left(\tildbm{\omega}^2\right)\,
\end{equation}
and
\begin{equation}
    \label{eq:J_exp}
    \begin{split}
J=\det\left(\bm{F}\right)=\det\left(\bm{F}_0\right)\det\left[\bm{I}+\bm{F}_0^{-1}\left(\bm{F}-\bm{F}_0\right)\right]\simeq J_0\left[1+\tr\left(\tildbm{\varepsilon}\right)+\frac{1}{2}\tr^2\left(\tildbm{\varepsilon}\right)-\frac{1}{2}\tr\left(\tildbm{\varepsilon}^2\right)-\frac{1}{2}\tr\left(\tildbm{\omega}^2\right)\right]\,,
\end{split}
\end{equation}
\noindent where we used \eqref{eq:tilde_eps}-\eqref{eq:tilde_omega}, $\tr\left(\tildbm{\varepsilon}^T\,\tildbm{\omega}\right)=\tr\left(\tildbm{\omega}^T\,\tildbm{\varepsilon}\right)=0$ along with $\det\left(\bm{\xi}\right)\simeq\tr\left(\bm{\xi}\right)+\,\tr^2(\bm{\xi})/2-\,\tr(\bm{\xi}^2)/2$ for $\bm{\xi}\simeq\bm{I}$. Also,
\begin{equation}
    \label{eq:logJ_exp}
    \begin{split}
\log(J)&\simeq\log(J_0)+\log\left[1+\tr\left(\tildbm{\varepsilon}\right)+\frac{1}{2}\tr^2\left(\tildbm{\varepsilon}\right)-\frac{1}{2}\tr\left(\tildbm{\varepsilon}^2\right)-\frac{1}{2}\tr\left(\tildbm{\omega}^2\right)\right]\simeq\log(J_0)+\tr\left(\tildbm{\varepsilon}\right)-\frac{1}{2}\tr\left(\tildbm{\varepsilon}^2\right)-\frac{1}{2}\tr\left(\tildbm{\omega}^2\right)\,,
\end{split}
\end{equation}
\noindent using $\log(1+\xi)\simeq \xi-\xi^2/2$ for $\xi\ll1$. From \eqref{eq:J_exp}  we also obtain
\begin{equation}
    \label{eq:Jtilde2_exp}
\left(J-J_0\right)^2\simeq J_0^2\tr^2\left( \tildbm{\varepsilon}\right)\,.
\end{equation}

\noindent The expansion of the entropic term $\Psi_{ent}$ reads
\begin{equation}
    \label{eq:ent_exp}
    \begin{split}
\frac{1}{2}Nk_BT\left[ \mathrm{tr}(\bm{C}) - 3 -2 \log(J) \right]\simeq \frac{1}{2}&Nk_BT\left[ 3\delta_0^2 -3 -2 \log(J_0) \right]+\\&+{Nk_BT}\left[\left(\delta_0^2-1\right)\,\tr\left(\tildbm{\varepsilon}\right)+\frac{1}{2}\left(\delta_0^2+1\right)\,\tr\left(\tildbm{\varepsilon}^2\right)-\frac{1}{2}\left(\delta_0^2-1\right)\,\tr\left(\tildbm{\omega}^2\right)\right]\,
\end{split}
\end{equation}
\noindent where we used \eqref{eq:C_exp} and  \eqref{eq:logJ_exp}. Expanding $\Psi_{bulk}$ leads to 
\begin{equation}
    \label{eq:bulk_exp}
    \begin{split}
\frac{K}{2}\left[J-\left(1+\Omega\Phi\right)\right]^2&=\frac{K}{2}J^2+\frac{K}{2}\left(1+\Omega\Phi\right)^2-{KJ}\left(1+\Omega\Phi\right)\simeq\\
&\simeq\frac{K}{2}J_0^2+KJ_0\left(J-J_0\right)+\frac{K}{2}\left(J-J_0\right)^2+\frac{K}{2}\left(1+\Omega\Phi_0\right)^2+K\Omega\left(1+\Omega\Phi_0\right)J_0\tilde c+\frac{K}{2}\Omega^2J_0^2\tilde c^2+\\
 &-KJ_0\left[1+\Omega\left(\Phi_0+J_0 \tilde c\right)\right]\left[1+\tr\left(\tildbm{\varepsilon}\right)+\frac{1}{2}\tr^2\left(\tildbm{\varepsilon}\right)-\frac{1}{2}\tr\left(\tildbm{\varepsilon}^2\right)-\frac{1}{2}\tr\left(\tildbm{\omega}^2\right)\right]\simeq\\
 &\simeq\frac{K}{2}\left[J_0-\left(1+\Omega\Phi_0\right)\right]^2-K\Omega\left[J_0-\left(1+\Omega\Phi_0\right)\right] J_0\tilde c+\frac{K}{2}\Omega^2 J_0^2 \tilde c^2+\\
& +{KJ_0}\left[J_0-\Omega J_0 \tilde c-\left(1+\Omega\Phi_0\right)\right]\tr\left(\tildbm{\varepsilon}\right)+\frac{KJ_0}{2}\left[2J_0-\Omega J_0\tilde c-\left(1+\Omega\Phi_0\right)\right]\tr^2\left(\tildbm{\varepsilon}\right)+\\
&-\frac{KJ_0}{2}\left[J_0-\Omega J_0 \tilde c-\left(1+\Omega\Phi_0\right)\right]\tr\left(\tildbm{\varepsilon}^2\right)-\frac{KJ_0}{2}\left[J_0-\Omega J_0\tilde c-\left(1+\Omega\Phi_0\right)\right]\tr\left(\tildbm{\omega}^2\right)
\,
\end{split}
\end{equation}
\noindent where we accounted for \eqref{eq:J_exp} and \eqref{eq:Jtilde2_exp}. With \eqref{eq:cauchy_stress_flory_int}, we can simplify \eqref{eq:bulk_exp} as 
\begin{equation}
    \label{eq:bulk_exp2}
    \begin{split}
\frac{K}{2}\left(J-\left(1+\Omega\Phi\right)\right)^2&\simeq \frac{K}{2}\left[J_0-\left(1+\Omega\Phi_0\right)\right]^2-K\Omega J_0\tilde c\left[J_0-\left(1+\Omega\Phi_0\right)\right]+\frac{K}{2}\Omega^2 J_0^2 \tilde c^2-Nk_BT\left(\delta_0^2-1\right)\tr\left(\tildbm{\varepsilon}\right)+\\
&-{KJ_0^2}\Omega\, \tilde c\,\tr\left(\tildbm{\varepsilon}\right)+\frac{KJ_0^2}{2}\Omega\,\tr^2\left(\tildbm{\varepsilon}\right)-\frac{Nk_bT}{2}\left(\delta_0^2-1\right)\tr^2\left(\tildbm{\varepsilon}\right)+\frac{Nk_bT}{2}\left(\delta_0^2-1\right)\tr\left(\tildbm{\varepsilon}^2\right)+\\
&+\frac{Nk_bT}{2}\left(\delta_0^2-1\right)\tr\left(\tildbm{\omega}^2\right)
\,.
\end{split}
\end{equation}
\noindent Summing \eqref{eq:ent_exp} and \eqref{eq:bulk_exp2} we obtain the expanded energy density
\begin{equation}
    \label{eq:bulk_exp3}
    \begin{split}
\Psi_{el}(\bm{F}, \Phi) &\simeq \Psi_{exp}\left(\tildbm{\varepsilon}, \tilde c\right) =\underbrace{\frac{1}{2}Nk_BT\left[3\delta_0^2 -3 -2 \log(J_0) \right]+\frac{K}{2}\left[J_0-\left(1+\Omega\Phi_0\right)\right]^2}_{\Psi_{0}} +\frac{K}{2}\Omega^2J_0^2\tilde c^2+\\
&-K\Omega J_0 \tilde c\left[J_0-\left(1+\Omega\Phi_0\right)\right] -{KJ_0^2}\Omega\tilde c\,\tr\left(\tildbm{\varepsilon}\right)+\frac{1}{2}\left[{KJ_0^2}-Nk_BT\left(\delta_0^2-1\right)\right]\tr^2\left(\tildbm{\varepsilon}\right)+Nk_BT\delta_0^2\,\tr\left(\tildbm{\varepsilon}^2\right)\,,
\end{split}
\end{equation}
\noindent where $\Psi_{0}$ is the elastic energy density in the intermediate configuration. 

\section{Computational  aspects} % ---------Computational aspects
\begin{figure}[!hbt]
    % \hspace*{-2cm}
    \centering\captionsetup{justification=centering}
    \includegraphics[scale = 0.4, trim={2cm 0 0 2cm}, clip]{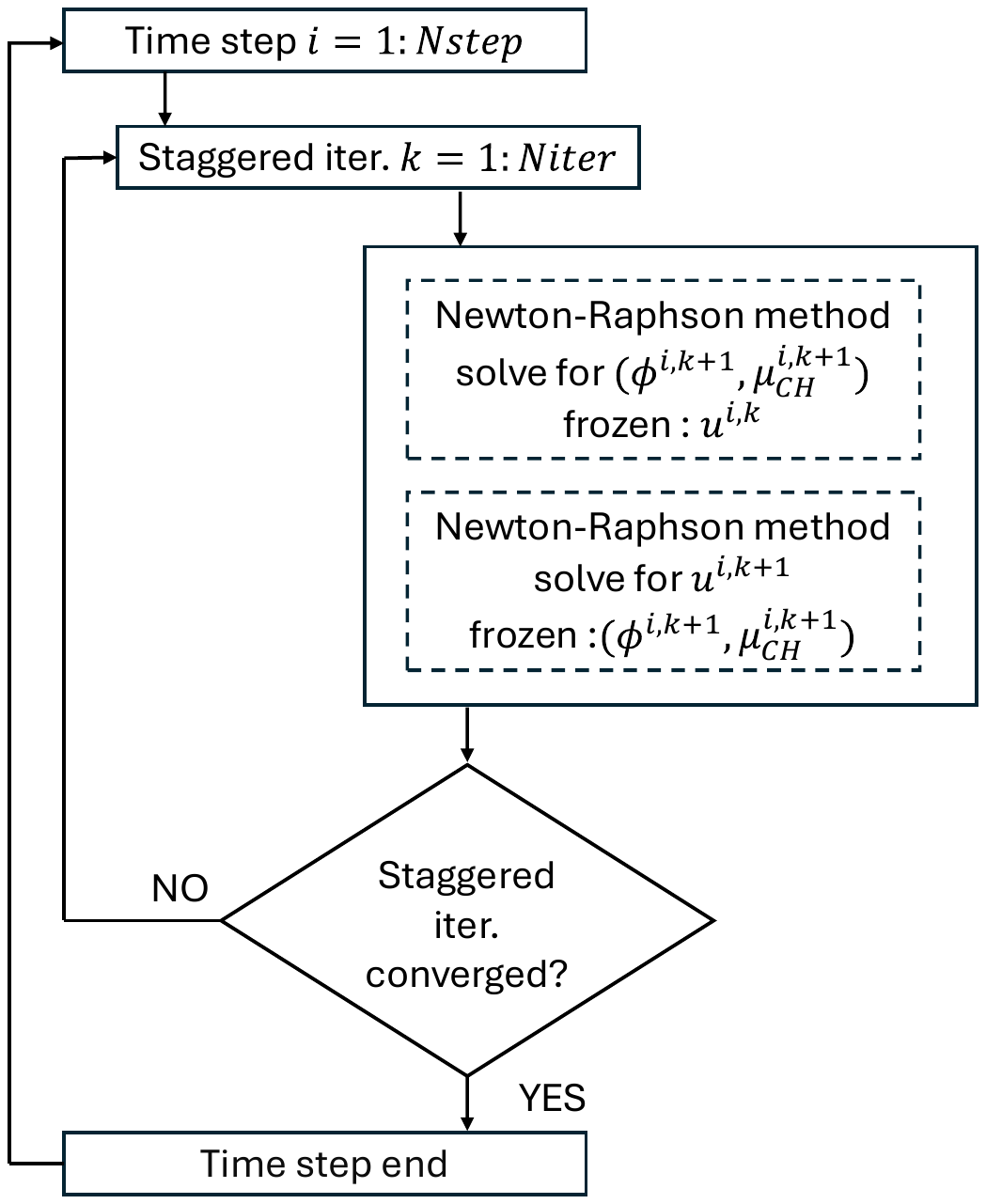}
    \caption{Staggered solution strategy.}
\label{fig:staggered_algorithm}
\end{figure}

	In this section we provide some computational details needed for the implementation of the model.
\label{app:solution_strategy} 
We perform time discretization with implicit finite differences and space discretization with linear (in 1D) or bilinear (in 2D) finite elements. 
The coupled governing equations (namely, \eqref{eq:1d_mass_balance_phi} and \eqref{eq:1d_linear_momentum} in 1D and \eqref{eq:2d_mass_balance_linear_mom} in 2D) are solved in a staggered fashion, which involves the alternate solution of the mechanical and of the mass balance equation until convergence is met (Figure \ref{fig:staggered_algorithm}). The fourth-order mass balance equation is  solved using the  mixed formulation proposed in \cite{elliott_numerical_1987, wodo_computationally_2011, kaessmair_comparative_2016, zhang_numerical_2021}. This involves the definition of the {Cahn-Hilliard} chemical potential $\tilde \mu_{CH}$ as an additional unknown field allowing to split the mass balance equation into two second-order partial differential equations. Hence, in the 2D case we have 
\begin{equation}
    \label{eq:2d_mixed_formulation}
    \left\{
    \begin{aligned}
        & \frac{\partial \phi}{\partial t}= M \left(\Delta_{\!x} \tilde\mu_{CH} - \alpha \phi \right), \\
        & \tilde \mu_{CH} = \gamma \left( \xi\phi + 4\beta\phi^3 \right) - \gamma \kappa \Delta_x \phi -m\,\text{tr}(\bm{\tilde \varepsilon})\,,
    \end{aligned}
    \right.
\end{equation}
\noindent which, in the 1D case, reduces to 
\begin{equation}
    \label{eq:1d_mixed_formulation}
    \left\{
    \begin{aligned}
        & \frac{\partial \phi}{\partial t}= M \left( \tilde\mu_{CH,xx} - \alpha \phi \right), \\
        & \tilde \mu_{CH} = \gamma \left( \xi\phi + 4\beta\phi^3 \right) - \gamma \kappa \phi_{,xx} - m\tilde{\varepsilon}\,.
    \end{aligned}
    \right.
\end{equation}
\noindent We then solve the mass balance equations for  $(\phi, \, \tilde\mu_{CH})$ and the mechanical problem for $u$ using a Newton-Raphson scheme up to a tolerance $tol_{NR}$. The staggered iterations among the two problems are performed until the $L_2$ norm of the residuals fall below a given tolerance $tol_{stag}$. For both 1D and 2D computations, the adopted tolerances are $tol_{NR} = 10^{-9}$ and $tol_{stag} = 10^{-7}$.

Efficient time integration of the {Cahn-Hilliard} or {Cahn-Oono} equations is still an open issue due to the different time scales involved in the phase separation dynamics \cite{ eberhardsteiner_isogeometric_2009, wodo_computationally_2011, li_computationally_2017, barros_finite_2021}. In particular, the emergence of the spinodal structure during the early stage of  phase separation and the dissolution of the interfaces during coarsening take place almost instantaneously, while the diffusive process between subsequent interface disappearence events can be two or three orders of magnitude slower. While choosing a constant large time step can lead to numerical instability problems and inconsistent results during the early stage, a constant small time step requires a very high number of time steps to capture the coarsening stage. To improve the computational efficiency, we adopt here a backward Euler scheme enhanced with a simple adaptive time-stepping scheme similar to what proposed in \cite{kuhl_computational_2007}. It consists in continuously adjusting the time step size depending on the number of staggered iterations required before convergence. We increase the time step by $0.1\%$ in the 1D model and by $1\%$ in the 2D model whenever the number of staggered iterations needed before convergence is less than or equal to ten and we halve it otherwise. Despite its heuristic nature, this approach manages to capture efficiently the different evolution stages in the computations illustrated here. 

\section{Calibration procedure} % --------- Calibration procedure---------------

We detail here the procedure adopted to calibrate the proposed model.

\subsection{Calibration of the chemical parameters $\gamma, \, \beta, \, \kappa$ in the 1D model} % ----------------- beta, gamma, kappa calibration ---------------------------
\label{app:1d_calibration}

The parameters of the Ginzburg-Landau free energy density are calibrated to capture the experimental observations. According to   \eqref{eq:1d_tot_energy}, the parameter $\gamma$ controls the relative importance of the elastic and the chemical driving forces during spinodal decomposition, with the mechanical contribution becoming negligible for $\gamma \gg E_0$. In this limit case, the initial characteristic length $\ell_0^u$ and time $\tau^u$ are the same of the standard Cahn-Hilliard uncoupled model and read
\begin{equation}
    \label{eq:1d_gamma_infty_ell0}
    \lim_{\gamma \gg E_0} \ell_0 = 2\pi \sqrt{\frac{-2\kappa}{\xi}}\,,\quad\text{and}\quad
    \lim_{\gamma \gg E_0} \tau = \frac{8\pi\kappa}{M\gamma\xi^2}\,
\end{equation}
\noindent On the other hand, the importance of the mechanical contribution increases when $\gamma \ll E_0$ leading to
\begin{equation}
    \label{eq:1d_gamma_zero_ell0_tau}
        \lim_{\gamma \ll E_0} \ell_0 =\ell^m_0= 2\pi\sqrt{ \frac{2\kappa}{\left(m^2/E_0 \right)}} \sqrt{\gamma}\,, \quad \text{and}\quad \lim_{\gamma \ll E_0} \tau = \tau^m=\frac{8\pi\kappa}{M\left(m^2/E_0 \right)^2} \gamma\,,
\end{equation}
\noindent These limit values are independent of the temperature parameter $\xi$ and charaterized by $\ell^m_0\propto \sqrt{\gamma}$ and $\tau^m\propto \gamma$. It is thus clear that, to predict an influence of both mechanical and chemical contributions, we should use intermediate values of $\gamma$.

\begin{figure}[h!]
    % \hspace*{-2cm}
    \centering
    \captionsetup{justification=centering}
    \includegraphics[scale = 0.8]{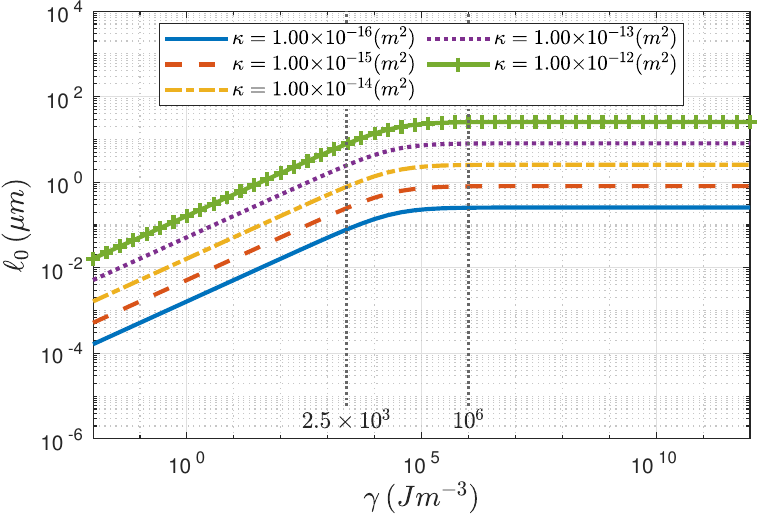}
    \caption{Influence of $(\gamma, \, \kappa)$ on the initial characteristic length $\ell_0$, with $350 \, (kPa)$ taken as a reference for the experimental parameters}
    \label{fig:gamma_kappa_ell0}
\end{figure}

Fig.\ref{fig:gamma_kappa_ell0} shows a bi-logarithmic plot of the $\ell_0$ vs. $\gamma$ curve \eqref{eq:1d_initial_characteristic_length} for different values of $\kappa$, while the other parameters are set equal to those of the material with $E_{PDMS}=350 \, kPa$. For $\gamma\leq 2.5\times10^3 \, Jm^{-3}$, the curve show a trend consistent with \eqref{eq:1d_gamma_zero_ell0_tau} and, hence, mainly influenced by the mechanical energy. Conversely, for $\gamma \geq 10^6 \, Jm^{-3}$, $\ell_0$ attains the uncoupled value $\ell_0^u$. Since the initial characteristic length observed experimentally varies with the elastic properties of the mixture, we choose $\gamma = 2.5\times10^3 \, Jm^{-3}$.

Also, Fig.\ref{fig:gamma_kappa_ell0} shows that the parameter $\kappa$  controls the initial characteristic length $\ell_0$. Hence, we tune $\kappa$ so that $\ell_0$ is equal to the experimental value associated to the $350 \, kPa$ PDMS. The chosen values of $\gamma$ and $\kappa$ are then used for all other types of PDMS. To our knowledge, there are no experimental data quantifying $\gamma, \kappa$ in the literature. The values selected here are comparable to those used in \cite{anders_isogeometric_2012, inguva_continuum-scale_2021, cabral_spinodal_2018} for similar polymer blends.

The parameter $\beta$ controls the amplitude of the two equilibrium phases, i.e. the binodal points $\phi_b^i, \, i = 1, \, 2$ and does not influence the initial characteristic length and time nor the coarsening stage. The binodals are determined by the \textit{Maxwell} common tangent equation\footnote{One approach to obtain the \textit{Maxwell} common tangent is to consider two phases separated by a sharp interface and apply the \textit{Coleman-Noll} procedure \cite{coleman_thermodynamics_nodate} to characterize the equilibrium of the system. Using balance of mass, mechanical equilibrium, rate of total energy and the \textit{Clausius–Duhem} inequality, we can derive the interfacial driving force at equilibrium which gives the \textit{Maxwell} tangent equation \eqref{eq:1d_binodals}. The adaptation of the \textit{Coleman-Noll} procedure to the case including phase separation and elasticity is detailed in \cite{guin_electromechanical_2018}, Chapt.~7. }
\begin{equation}
    \label{eq:1d_binodals}
    \begin{split}
        \mu \left( \phi_{b}^1 \right) = \mu \left( \phi_{b}^2 \right) & = \frac{ [\![ \psi_b]\!] + [\![ \psi_{el}]\!] - \langle {\sigma} \rangle \cdot  [\![ {\varepsilon} ]\!]}{[\![ \phi ]\!]}, 
    \end{split}
\end{equation}
\noindent where $[\![\bullet]\!]=\bullet(\phi_{b}^2)-\bullet(\phi_{b}^1)$ and $\langle {\bullet} \rangle=0.5[\bullet(\phi_{b}^2)-\bullet(\phi_{b}^1)]$ are the jump and the average of the field $(\bullet)$ across the interface. Eq. \eqref{eq:1d_strain} provides the relation between the strain and the phase-field variable, while for our stress-free domain in the 1D model the last term vanishes. Therefore we obtain
\begin{equation}
    \label{eq:1d_binodals_coupled}
    \phi_{b}^1 = \sqrt{ -\frac{\xi -m^2/(E_0\gamma)}{4\beta}  },  \quad \phi_{b}^2 = -\sqrt{ -\frac{\xi -m^2/(E_0\gamma)}{4\beta}  }
\end{equation}
\noindent In the uncoupled situation where $[\![ \psi_{el}]\!] = 0$,  \eqref{eq:1d_binodals} simplifies to
\begin{equation}
    \label{eq:1d_binodals_uncoupled}
    \phi_{b}^1 = \sqrt{ \frac{-\xi}{4\beta}  },  \quad \phi_{b}^2 = -\sqrt{ \frac{-\xi}{4\beta}  }
\end{equation}

\begin{figure}[h!]
    \hspace*{-1cm}
    \centering
    \captionsetup{justification=centering}
    \includegraphics[scale = 0.7]{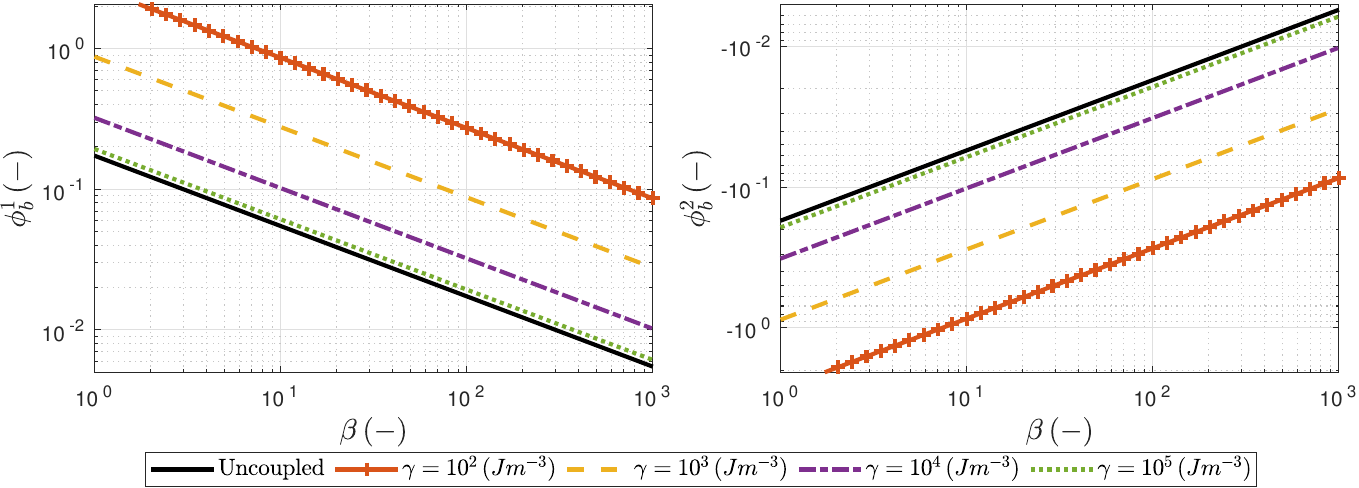}
    \caption{Influence of $\gamma$ and $\, \beta$ on the amplitudes of the binodals for the material with  $E_{PDMS}=350 \, (kPa)$.}
    \label{fig:1d_influence_beta_gamma_binodals}
\end{figure}

Fig.~\ref{fig:1d_influence_beta_gamma_binodals} illustrates the curves relating binodal points and $\beta$ for different values of $\gamma$ and their comparison with the uncoupled case. For large values of $\gamma \gg E_0$, the uncoupled behavior is approached, while for $\gamma \ll E_0$, the binodals are mainly controlled by $m$, i.e. by the {osmotic} contribution. Experimentally, no measurement of $\phi_{b}^i$ was conducted. Yet, bright-field optical microscopy analyses \cite{fernandez-rico_elastic_2023} suggest that the binodals are very close to the initial homogeneous value $\phi_0=0$. Consequently, we set $\beta=\beta^*$ such that $\phi_{b}^i \sim \phi_0 \pm 0.1$, which represents a volume fraction variation of about $\varphi_0\pm 5\%$. Using \eqref{eq:1d_binodals_coupled} we obtain
\begin{equation}
    \label{eq:1d_beta_star}
    \beta^* = -25\left(\xi - \frac{m^2}{E_0\gamma}\right)\,.
\end{equation}
\noindent 

\subsection{Estimation of the expansion coefficient $\Omega$} % ------ expansion coefficient estimation -------
\label{app:swelling_coefficient}

From \cite{fernandez-rico_elastic_2023} we have the swelling of the PDMS during incubation $r_0$ expressed in the reference configuration as ratio between mass increase due to HFBMA oil intake $m_{HFBMA}$ and mass of dry polymer $m_{PDMS}$
\begin{equation}
    r_0 = \frac{m_{HFBMA}}{m_{PDMS}}\,.
\end{equation}  
\noindent Assuming that the increase of volume  during swelling is equal to the volume of the absorbed oil, the volumetric swelling ratio $s_o$ reads
\begin{equation}
    s_0 = \frac{\rho_{PDMS}}{\rho_{HFBMA}}r_0\,,
\end{equation}
\noindent where $\rho_{HFBMA}$ and $\rho_{PDMS}$ are the densities of the HFBMA oil and of the dry PDMS, respectively. The swelling ratio is related to the deformation gradient $\bm{F}_0 = \delta_0 \bm{I}$ by
\begin{equation}
    \mathrm{det}(\bm{F}_0) = J_0 = 1 + s_0\,.
\end{equation}
\noindent Hence, the initial stretch $\delta_0$ is
\begin{equation}
    \label{app:eq_stretch_expressions}
    \delta_0 = 
    \begin{cases}
        1+s_0, & \text{in 1D}\\
        \left(1+s_0\right) ^{1/2}, & \text{in 2D} \\
        \left(1+s_0\right) ^{1/3}, & \text{in 3D}\,.
    \end{cases}
\end{equation}

 The swelling due to a variation of the volume fraction of the solvent can be expressed as
\begin{equation}
    \Omega \varphi \bm{I}  = \frac{1}{2} \Omega \phi + \bm{e}_0\,,
\end{equation}
\noindent where $\bm{e}_0 = \Omega \varphi_0 \bm{I}$ is the Euler-Almansi strain related to $\varphi_0$ in the intermediate configuration. Considering the initial stretch \eqref{app:eq_stretch_expressions}, $\bm{e}_0$ takes the form
\begin{equation}
    \label{app:eq_euler_almansi}
    \bm{e}_0 = \frac{1}{2}\left( \bm{I} - \bm{F}_0^{-T}\bm{F}_0^{-1} \right) = \frac{1}{2}\left( 1 - \frac{1}{\delta_0^2} \right) \bm{I}\,.
\end{equation} 
\noindent Combining \eqref{app:eq_stretch_expressions} and \eqref{app:eq_euler_almansi} we have
\begin{equation}
    \Omega = 
    \begin{cases}
        \displaystyle\frac{1}{2\varphi_0}\left[1 - \frac{1}{\left( 1+s_0 \right)^2} \right], & \text{in 1D}\\
        \displaystyle\frac{1}{2\varphi_0}\left[1 - \frac{1}{\left( 1+s_0 \right)} \right], & \text{in 2D}\\
        \displaystyle\frac{1}{2\varphi_0}\left[1 - \frac{1}{\left( 1+s_0 \right)^{2/3}} \right], & \text{in 3D}\\
    \end{cases}
\end{equation}
\noindent Tab.~\ref{tab:table_Omega_exp_data} summarizes the experimental values used to estimate $\Omega$.

\begin{table}[h!]
    \centering
    \begin{tabular}{ccccccc}
$E_{PDMS}$ & $\varphi_0$  & $r_0$    \\
$(kPa)$    & $(-)$        & $(-)$    \\
$180$      & $0.65$       & $0.62$   \\
$350$      & $0.62$       & $0.61$   \\
$800$      & $0.56$       & $0.56$          
    \end{tabular}
    \caption{\label{tab:table_Omega_exp_data}Summary of experimental results necessary for the estimation of $\Omega$.}
\end{table}
\noindent with $\rho_{HFBMA} = 1.37 \, 10^{3} \, (kg \, m^{-3})$ and  $\rho_{PDMS} = 9.7 \, 10^{2} \, (kg \, m^{-3})$.

\subsection{Estimation of the coupling parameter $m$ in the 2D model} % --- m estimation --------
\label{app:2d_porous_coefficient}

Here we retrieve a 2D estimation of the osmotic coefficient $m$ as a function of the expansion coefficient per unit volume fraction $\Omega$.  Since the applied boundary conditions lead to a vanishing average stress, we assume the local stresses to be close to zero. Therefore, using \eqref{eq:stress_strain_2d} we have
\begin{equation}
    \begin{split}
        \tildbm{\sigma} \sim \bm{0} & \Rightarrow \tildbm{\varepsilon} \sim  \frac{m}{2(\lambda_0+G_0)}\phi \bm{I} \\
        & \Rightarrow \tildbm{\varepsilon} \sim \frac{m}{(\lambda_0+G_0)}\tilde{\varphi} \bm{I} \\
        & \Rightarrow \tildbm{\varepsilon} \sim \Omega\tilde{\varphi}\bm{I}\,,
    \end{split}
\end{equation}
\noindent finally delivering \eqref{eq:m_coefficient_2D}.

%%%%%%%%%%%%%%%%%%%%% Call the bibliography %%%%%%%%%%%%%%%
\bibliographystyle{elsarticle-num} 
\bibliography{Paper_Spinodal_decomposition}
\end{document}